\documentstyle[12pt,epsfig,axodraw]{article}

\newcommand{\lsim}{\raisebox{-0.13cm}{~\shortstack{$<$ \\[-0.07cm] $\sim$}}~}
\newcommand{\gsim}{\raisebox{-0.13cm}{~\shortstack{$>$ \\[-0.07cm] $\sim$}}~}

\newcommand{\ra}{\rightarrow}

\newcommand{\s}{\\ \vspace*{-3mm} }
\newcommand{\nn}{\noindent}
\newcommand{\non}{\nonumber}
\newcommand{\beq}{\begin{eqnarray}}
\newcommand{\eeq}{\end{eqnarray}}
\newcommand{\tb}{\tan\beta}
\newcommand{\mx}{m_{\tilde{\chi}_1^0}}
\newcommand{\msqa}{m_{\tilde{f}_1}}
\newcommand{\msqb}{m_{\tilde{f}_2}}
\newcommand{\sqa}{\tilde{f}_1}
\newcommand{\sqb}{\tilde{f}_2}
\newcommand{\sq}{s_{2\theta_{\tilde{f}}}}
\newcommand{\cq}{c_{2\theta_{\tilde{f}}}}
\newcommand{\lsp}{\mbox{$\tilde\chi_1^0$}}

\begin{document}

\begin{flushright}
PM/01--28\\
TUM-HEP-434-01 \\
MPI-TH-32-01 \\
September 2001
\end{flushright}

\vspace{1cm}

\begin{center}

{\large {\bf \mbox{Loop induced Higgs and Z boson couplings to Neutralinos}}}

\vspace*{3mm}

{\large {\bf and implications for collider and Dark Matter searches}} 

\vspace{1cm}

{\sc A. Djouadi$^1$, M. Drees$^2$, P. Fileviez Perez$^3$} and {\sc
M. M\"uhlleitner$^1$}

\vspace*{0.5cm}

$^1${\it Laboratoire de Physique Math\'ematique et Th\'eorique,
UMR5825--CNRS,
\\ Universit\'e de Montpellier II, F--34095 Montpellier Cedex 5, France.}

\vspace*{3mm}

$^2${\it Physik Department, Technische Universit\"at M\"unchen, \\ 
James Franck Strasse, D--85748 Garching, Germany.}

\vspace*{3mm}

$^3${\it Max-Planck Institut f\"ur Physik  (Werner-Heisenberg Institut)
\\ F\"ohringer Ring 6, 80805 M\"unchen, Germany.}

\end{center}

\vspace{1cm}

\begin{abstract}

\nn 
We calculate the one-loop induced couplings of two gaugino--like
neutralinos to the $Z$ and Higgs bosons in the Minimal Supersymmetric
Standard Model. These couplings, which vanish at the tree level, can
be generated through loops involving fermions and sfermions. We show
that, while the neutralino contribution to the invisible $Z$ boson
decay width remains small, the loop induced couplings to the lightest
Higgs boson might be sufficiently large to yield a rate of invisible
decays of this Higgs boson that should be detectable at future
$e^+e^-$ colliders. We also study the implications of these couplings
for direct searches of Dark Matter and show that they can modify
appreciably the neutralino--nucleon elastic cross section for some
parameter range.
\end{abstract}

\newpage

\section{Introduction}

The Minimal Supersymmetric Standard Model (MSSM) \cite{MSSM} is at the
moment considered to be the most plausible extension of the Standard
Model (SM). Out of the plethora of new particles contained in this
model the lightest neutralino \lsp\ plays a special role. It is often
the lightest of all superparticles (LSP), which is absolutely stable
if R--parity is conserved. This means that neutralino LSPs produced
at colliders will be invisible, leading to the famous ``missing
[transverse] momentum'' signatures for the production of
superparticles. In particular, a final state consisting of two LSPs
only would be entirely invisible, so that \lsp\lsp\ final states can
contribute to the invisible width of the $Z$ or neutral Higgs
bosons. Moreover, if the LSP is stable, we expect some relic LSPs from
the Big Bang era to still exist today, in addition to the well--known
relic neutrinos and (microwave) photons. In fact, it was realized
almost twenty years ago that \lsp\ is a good (cold) Dark Matter
(DM) candidate \cite{NeutDM}. The LSP--nucleon scattering cross
section, which determines the size of the expected signal in direct DM
detection experiments \cite{SDM}, depends on the size of the LSP
couplings to the $Z$ and Higgs bosons; these couplings can also play a
role in the calculation of the \lsp\ relic density. A precise knowledge
of these couplings is therefore important for both collider phenomenology
and cosmology. \s

The lightest neutralino \lsp\ couples to the $Z$ boson and to the MSSM
Higgs bosons only if it has a non--vanishing higgsino component. On
the other hand, most models predict \lsp\ to be dominantly a gaugino,
in particular, a bino-- or photino--like state. Moreover, an LSP with
dominant higgsino component has a thermal relic density of the
required magnitude only if its mass is in the TeV range, beyond the
reach of near--future colliders and also beyond the range of masses
that is usually considered to be natural. In contrast, a bino--like
LSP is a good thermal DM candidate for masses in the 100 GeV range;
masses in this range are more natural, and can be probed at present
and near--future colliders. Since these gaugino--like states have
suppressed tree--level couplings to gauge and Higgs bosons, loop
contributions to these couplings might be significant. \s

In this paper we compute one-loop corrections to the neutralino
couplings to the $Z$ and Higgs bosons, where we only consider
contributions with fermions and sfermions inside the loop [the other
possible contributions, involving charginos, neutralinos and Higgs or
gauge bosons, vanish in the bino--like limit for the \lsp\ ]. Our
analytical results for these corrections, for arbitrary momenta and
general $\tilde f_L - \tilde f_R$ mixing\footnote{Loop corrections to
the $Z$ exchange contribution to the bino annihilation cross section
have previously been computed in \cite{sred}. However, in this earlier
paper $\tilde f_L - \tilde f_R$ mixing has been ignored, and
analytical results are only given for vanishing LSP
three--momenta. Our numerical results agree qualitatively with
their's.}, are given in Sec.~2. In Sec.~3 we apply these results to
compute the loop--corrected invisible branching ratios of $Z$ and
Higgs bosons, and in Sec.~4 we study the loop--corrected LSP--nucleon
scattering rate and relic density. \s

For the neutralino couplings to Higgs and gauge bosons, we found
in all cases corrections of up to a factor of two for reasonable values of 
the input parameters. The contribution of bino--like LSPs to the invisible
decay width of the lightest MSSM Higgs boson might be measurable at future 
high--energy and high--luminosity $e^+e^-$ colliders, while the contribution
to the invisible decay width of the $Z$ boson remains several orders of 
magnitude
below the present bound. On the other hand, the loop corrections can modify
appreciably the LSP--nucleon scattering cross section in some MSSM parameter
range.

\section{Neutralino couplings to Higgs and Z Bosons}

\subsection{Tree-level couplings} 

At the tree level, the couplings of the neutralinos $\tilde{\chi}_i^0$
to the $Z$ boson are given by the vertex\footnote{We assume
that all soft breaking parameters as well as $\mu$ are real,
i.e. conserve CP. We can then work with a real, orthogonal neutralino
mixing matrix $N$ if we allow the eigenvalues $m_{\tilde \chi_i^0}$ to
be negative. Note also that the vertex factors $\Gamma$ given in the
text are $2i$ times the coefficients of the relevant terms in the
interaction Lagrangian.} \cite{HK}:
\beq \label{zcoup0}
\Gamma^0_\mu(Z\tilde{\chi}_i^0\tilde{\chi}_j^0) = i\frac{g}{2c_W}\; 
\gamma_\mu \gamma_5 \; [N_{i3} N_{j3}-N_{i4}N_{j4} ] 
\equiv i g^0_{Z \tilde \chi_i^0 \tilde \chi_j^0} \gamma_\mu \gamma_5,
\eeq
while the neutralino couplings to the neutral CP--even Higgs bosons $\phi=h,H$ 
and to the CP--odd boson $A$ read \cite{HHG,R1}: 
\beq \label{hcoup0}
\Gamma^0(\phi\tilde{\chi}_i^0\tilde{\chi}_j^0) &=& i \frac{g}{2}  
\left[ (N_{i2} - \tan\theta_W N_{i1} ) 
( d_\phi N_{j3} + e_\phi N_{j4} ) \ + \ i \leftrightarrow j \right]
\equiv i g^0_{\phi \tilde \chi_i^0 \tilde \chi_j^0}; \\
\label{acoup0}
\Gamma^0(A\tilde{\chi}_i^0\tilde{\chi}_j^0) &=& \frac{g}{2}
\left[ (N_{i2} - \tan\theta_W N_{i1})
(d_A N_{j3} +e_A N_{j4}) \ + \ i \leftrightarrow j \right] \gamma_5 
\equiv g^0_{A \tilde \chi_i^0 \tilde \chi_j^0} \gamma_5.
\eeq
Here, $g=e/s_W$ is the SU(2) gauge coupling with $s_W^2=1-c_W^2 \equiv
\sin^2 \theta_W$. The quantities $d_\Phi, e_\Phi \ (\Phi = h, H, A)$
are determined by the ratio $\tan\beta$ of the vacuum expectation
values of the two doublet Higgs fields which are needed to break the
electroweak symmetry in the MSSM, and the mixing angle $\alpha$ in the
CP--even neutral Higgs sector:
\beq \label{deh}
d_H &=& - \cos\alpha \ , \ d_h = \sin\alpha \ , \ d_A= \sin \beta, \nonumber\\
e_H &=& \sin\alpha \ , \  e_h = \cos\alpha \ , \  e_A= -\cos \beta . 
\eeq
$N$ is the matrix which diagonalizes the four dimensional neutralino mass 
matrix:
\begin{eqnarray} \label{matrix}
{\cal M}_N = \left[ \begin{array}{cccc}
M_1 & 0 & -m_Z s_W c_\beta & m_Z  s_W s_\beta \\
0   & M_2 & m_Z c_W c_\beta & -m_Z  c_W s_\beta \\
-m_Z s_W c_\beta & m_Z  c_W c_\beta & 0 & -\mu \\
m_Z s_W s_\beta & -m_Z  c_W s_\beta & -\mu & 0
\end{array} \right] ,
\end{eqnarray}
where $M_1$ and $M_2$ are the SUSY breaking masses for the U(1)$_Y$
and SU(2)$_L$ gauginos, $\mu$ is the higgsino mass parameter, and
$s_\beta \equiv \sin \beta$, etc. This matrix can be diagonalized
analytically \cite{R2}, but the expressions of the neutralino masses
and the $N_{ij}$ matrix elements are rather involved. However, if the
entries in the off--diagonal $2 \times 2$ submatrices in
eq.~(\ref{matrix}) are small compared to (differences of) the diagonal
entries, one can expand the eigenvalues in powers of $m_Z$
\cite{MD1,R3}:
\begin{eqnarray} \label{massapp}
m_{\tilde{\chi}_{1}^0} &\simeq& 
M_1 - \frac{m_Z^2}{\mu^2-M_1^2} \left( M_1 +\mu s_{2\beta} \right) s_W^2;
 \non \\
m_{\tilde{\chi}_{2}^0} &\simeq& 
M_2 - \frac{m_Z^2}{\mu^2-M_2^2} \left( M_2 +\mu s_{2\beta} \right) c_W^2;
 \non \\
m_{\tilde{\chi}_{3}^0} &\simeq& 
-\mu - \frac{m_Z^2(1- s_{2\beta})} {2} \left( \frac{s_W^2} {\mu+M_1}
+ \frac{c_W^2} {\mu+M_2} \right); \non \\
m_{\tilde{\chi}_{4}^0} &\simeq& 
\mu + \frac{m_Z^2(1+ s_{2\beta})} {2} \left( \frac{s_W^2} {\mu-M_1}
+ \frac{c_W^2} {\mu-M_2} \right). 
\end{eqnarray}
In this analysis, we are interested in the situation $|\mu| > M_1, \ M_2$ and
$\mu^2 \gg m_Z^2$. In this case the lighter two neutralinos will be
gaugino--like. If $|M_1| < |M_2|$, which is the case if gaugino masses
unify at the same scale where the gauge couplings appear to meet
\cite{gaugeuni}, the lightest state will be bino--like, and the 
next--to--lightest state will be wino--like. The two heaviest states
will be dominated by their higgsino components.\footnote{Loop corrections can
significantly change the mass splitting between the two higgsino--like
states \cite{giudice,higgsino}.} The eigenvectors of the mass matrix
(\ref{matrix}) can also be expanded in powers of $m_Z$. We find for
the bino--like state \cite{MD1,R3}:
\begin{eqnarray} \label{binostate}
N_{11} &=& \left[ 1+ \left( N_{12}/N_{11} \right)^2+
\left( N_{13}/N_{11} \right)^2+
\left( N_{14}/N_{11} \right)^2 \right]^{-1/2} ;
\non \\
\frac{N_{12}}{N_{11}} &=& \frac{m_Z^2 s_W c_W} {\mu^2 - M_1^2}
\frac { s_{2\beta} \mu + M_1 } {M_1-M_2} +{\cal O}(m_Z^3) ;
 \non \\
\frac{N_{13}}{N_{11}} &=&  m_Z s_W  
\frac {s_\beta \mu + c_\beta M_1}{\mu^2 - M_1^2} + 
{\cal O}(m_Z^2); \non \\
\frac{N_{14}}{N_{11}} &=&  - m_Z s_W  \frac{c_\beta \mu + s_\beta M_1}
{\mu^2 - M_1^2} 
+ {\cal O}(m_Z^2).
\end{eqnarray}
The corresponding expressions for the wino--like state read:
\begin{eqnarray} \label{winostate}
N_{22} &=& \left[ 1+ \left( N_{21}/N_{22} \right)^2+
\left( N_{23}/N_{22} \right)^2+
\left( N_{24}/N_{22} \right)^2 \right]^{-1/2} ;
\non \\
\frac{N_{21}}{N_{22}} &=& \frac{m_Z^2 s_W c_W} {\mu^2 - M_2^2}
\frac {s_{2\beta} \mu + M_2} {M_2-M_1} +{\cal O}(m_Z^3) ;
 \non \\
\frac{N_{23}}{N_{22}} &=&  -m_Z c_W  
\frac {s_\beta \mu + c_\beta M_2}{\mu^2 - M_2^2} + 
{\cal O}(m_Z^2); \non \\
\frac{N_{24}}{N_{22}} &=&  m_Z c_W  \frac{c_\beta \mu + s_\beta M_2}
{\mu^2 - M_2^2} 
+ {\cal O}(m_Z^2).
\end{eqnarray}
Note that the higgsino components of the gaugino--like states start at ${\cal
O}(m_Z)$, whereas the masses of these states deviate from their $|\mu|
\rightarrow \infty$ limit ($M_1$ and $M_2$) only at ${\cal O}(m_Z^2)$. \s

Inserting eqs.~(\ref{binostate}) into eq.~(\ref{zcoup0}) one sees that
the coupling of the LSP neutralinos to the $Z$ boson only occurs at
order $m_Z^2$, while eqs.~(\ref{hcoup0}) and (\ref{acoup0}) show that
the LSP couplings to the Higgs bosons $\Phi=h,H,A$ already receive
contributions at ${\cal O}(m_Z)$:
\beq \label{gzerobino}
g^0 (Z\tilde{\chi}_1^0\tilde{\chi}_1^0) & \sim & N_{13}^2- N_{14}^2 \ \sim \  
- s_W^2 c_{2\beta} \, \frac{m_Z^2}{\mu^2-M_1^2}; \non \\
g^0 (\Phi\tilde{\chi}_1^0\tilde{\chi}_1^0) &\sim & 
d_\Phi N_{13} + e_\Phi N_{14} \non \\
 &\sim&  s_W  m_Z \left[ \frac{(d_\Phi s_\beta - e_\Phi c_\beta) \mu}
{\mu^2 - M_1^2} + \frac{ ( d_\Phi c_\beta - e_\Phi s_\beta ) M_1 } 
{\mu^2 - M_1^2} \right] ,
\eeq
Similar expressions can be given for the couplings of the wino--like
state. This suppression of the tree--level couplings follows from the
fact that, in the neutralino sector, the $Z$ only couples to two
higgsino current eigenstates while a Higgs boson couples to one
higgsino and one gaugino current eigenstate, together with the fact
that mixing between current eigenstates is suppressed if $|\mu| \gg
m_Z$. Both kinds of couplings thus {\em vanish} as $|\mu| \rightarrow
\infty$. \s

The situation in the chargino sector is somewhat similar to what has been
discussed so far. The diagonalization of the chargino mass matrix
\begin{eqnarray} \label{charmat}
{\cal M}_C = \left[ \begin{array}{cc} M_2 & \sqrt{2}m_W s_\beta
\\ \sqrt{2}m_W c_\beta & \mu \end{array} \right]
\end{eqnarray}
leads, when expanded in powers of $m_W$, to the two chargino masses:
\begin{eqnarray}
m_{\tilde{\chi}_{1}^\pm}  \simeq   M_2 - \frac{m_W^2}{\mu^2-M_2^2} 
\left( M_2 +\mu s_{2\beta} \right) \simeq m_{\tilde \chi_2^0}\ \ , \ \ 
m_{\tilde{\chi}_{2}^\pm}  \simeq  \mu + 
\frac{m_W^2}{\mu^2-M_2^2} \left( M_2 s_{2 \beta} +\mu \right) ,
\end{eqnarray}
so that for $|\mu| \ra \infty$, the lightest chargino corresponds to a pure
wino state while the heavier chargino corresponds to a pure higgsino state. 
The couplings of the neutral Higgs bosons to chargino pairs $g^0(\Phi
\tilde{\chi}_i^\pm \tilde{\chi}_i^\mp)$ are suppressed in this limit, and 
only the couplings $g^0(\Phi \tilde{\chi}_1^\pm \tilde{\chi}_2^\mp)$
survive. Moreover, neither the $W$ nor the charged Higgs bosons
couple the LSP to the lighter chargino in this limit, $g^0(H^\pm
\tilde{\chi}_1^0 \tilde{\chi}_{1}^\mp) \sim {\cal O}(m_W/\mu), \ 
g^0(W^\pm \tilde{\chi}_1^0 \tilde{\chi}_{1}^\mp) \sim {\cal
O}(m^2_W/\mu^2)$. \s

However, the $Z$ boson does have full--strength couplings to pairs of
charginos $\tilde{\chi}^\pm_i \tilde{\chi}_i^\mp$, and LEP2 limits
imply that even the lighter chargino is too heavy to be produced in
the decay of the lighter neutral Higgs boson $h$. Moreover, the heavy
Higgs bosons can always undergo unsuppressed decays into at least some
SM fermions; in fact, decays involving $b-$quarks are usually enhanced
at large $\tb$. We therefore do not expect heavy Higgs decays into
modes that vanish for $|\mu| \rightarrow \infty$ to be significant
even after loop corrections have been applied. For these reasons we
will not discuss the chargino sector any further in this paper.

\subsection{One-loop induced couplings} 

At the one--loop level, the couplings of the lightest neutralinos to
the $Z$ and Higgs bosons can be generated, in principle, by diagrams
with the exchange of either sfermions and fermions, or of charginos or
neutralinos together with gauge or Higgs bosons, in the loop. However,
the latter class of diagrams can contribute to the couplings of Higgs bosons
to neutralinos only if one of the particles in the loop is a
higgsino. Similarly, these gauge--Higgs loops will contribute to the
coupling of bino--like LSPs to the $Z$ boson only if at least one
particle in the loop is a higgsino--like state. These loop
contributions will thus be suppressed by inverse powers of
$|\mu|$, in addition to the usual loop suppression factor. We
therefore do not expect them to be able to compete with the
tree--level couplings that exist for finite $|\mu|$, see
eq.~(\ref{gzerobino}).\footnote{Loops containing charged gauge bosons
and gauginos only can contribute to the $Z \lsp \lsp$ coupling if
\lsp\ is wino--like. However, these contributions are separately gauge
invariant only if the $Z$ boson is on--shell. Since in this scenario
$m_{\tilde \chi^\pm_1} \simeq m_{\tilde \chi^0_1}$, LEP searches imply
that \lsp\ is too heavy to be produced in the decays of on--shell $Z$
bosons. If the $Z$ boson is off--shell, box--diagrams with two charged
gauge bosons in the loop have to be added to obtain a gauge--invariant
result.} \s

We thus only consider diagrams with fermions and sfermions in the
loop, as shown in Fig.~1. In the case of the $Z \tilde{\chi}_1^0
\tilde{\chi}_1^0$ coupling, all three generations of matter particles
will be involved since they have full gauge coupling strength. In the
case of the $\Phi \tilde{\chi}_1^0 \tilde{\chi}_1^0$ couplings, only
the third generation (s)particles, which have large Yukawa couplings,
can give significant contributions to the amplitudes. Note that in the
bino limit there is no wave function renormalization to perform, since
the tree--level couplings are zero. Off--diagonal wave function
renormalization diagrams could convert one of the gaugino--like
neutralinos into a higgsino--like state, but this kind of contribution
is again suppressed by $1/|\mu|$, and can thus not compete with the
tree--level coupling. In case of the $Z$ coupling, both external
gauginos would have to be converted to higgsino--like states, which is
obviously only possible at the two--loop level. \s

\vspace*{-.5cm}
\begin{picture}(1000,200)(70,0)
\DashArrowLine(100,100)(160,100){4}
\Text(120,110)[]{$\Phi,Z$}
\ArrowLine(200,140)(160,100)
\ArrowLine(160,100)(200,60)
\DashArrowLine(200,60)(200,140){4}{}
\ArrowLine(250,140)(200,140)
\ArrowLine(200,60)(250,60)
\Text(260,130)[]{$\tilde\chi_1^0 (p_1)$}
\Text(260,70)[]{$\tilde\chi_1^0 (p_2)$}
\Text(210,100)[]{$\tilde{f}_i$}
\Text(180,130)[]{$f$}
\Text(180,70)[]{$f$}
\DashArrowLine(300,100)(360,100){4}
\Text(320,110)[]{$\Phi,Z$}
\DashArrowLine(360,100)(400,140){4}{}
\DashArrowLine(400,60)(360,100){4}{}
\ArrowLine(400,140)(400,60)
\ArrowLine(450,140)(400,140)
\ArrowLine(400,60)(450,60)
\Text(460,130)[]{$\tilde\chi_1^0(p_1)$}
\Text(460,70)[]{$\tilde\chi_1^0 (p_2)$}
\Text(100,140)[]{${\bf a)}$}
\Text(300,140)[]{${\bf b)}$}
\Text(410,100)[]{$f$}
\Text(380,130)[]{$\tilde{f}_i$}
\Text(380,70)[]{$\tilde{f}_j$}
\end{picture}
\vspace*{-1.5cm}

\nn Figure 1: The Feynman diagrams contributing to the one--loop couplings
of the lightest neutralinos to the $Z$ and $\Phi=h,H,A$ Higgs
bosons. Diagrams with crossed neutralino lines have to be added.
\setcounter{figure}{1}
\\ \\

We have calculated the contributions of these diagrams for arbitrary
momentum square of the Higgs and $Z$ bosons, finite masses for the
internal fermions and sfermions as well as for the external LSP
neutralinos, and taking into account the full mixing in the sfermion
sector. The amplitudes are ultra--violet finite as it should be. The
contributions from diagrams a) and b) to the $\Phi \tilde{\chi}_1^0
\tilde{\chi}_1^0$ couplings are separately finite for each fermion
species.\footnote{The contribution of diagram a) is finite only after
summation over both sfermion mass eigenstates.} The $Z
\tilde{\chi}_1^0 \tilde{\chi}_1^0$ couplings are finite once
contributions from both sfermion mass eigenstates of a given flavor
and from both diagrams in Fig.~1 are added; diagrams a) and b) are
separately finite after summation over a complete generation of
(s)fermions.  We have performed the calculation in the dimensional
reduction scheme \cite{R4}; since the one--loop couplings are finite
and do not require any renormalization, the same result is obtained
using the dimensional regularization scheme \cite{R5}, once the
contributions from a complete SU(2) doublet have been
added.\footnote{In the $\overline{\rm MS}$ scheme the contribution
from a given fermion species contains mass--independent terms
proportional to some combinations of $Z f f$ couplings. These constant
terms add to zero when summed over a complete generation of fermions
and sfermions.} The results are given below for a general gaugino
($|\mu| \gg M_1, M_2$, for arbitrary ordering of $M_1$ and $M_2$).

The one--loop induced $Z$ boson vertex to a pair of lightest 
neutralinos is given by ($p_1$ is an incoming, $p_2$ an outgoing momentum):
\beq \label{gonez}
\Gamma^1_\mu(Z\tilde{\chi}_1^0 \tilde{\chi}_1^0) &=& \frac{i}{4\pi^2} 
\frac{g}{c_W} \Bigg\{ \gamma_\mu \gamma_5 \sum_{f} N_c^{(f)}
\delta_a^{(f)} + (p_1-p_2)_\mu \gamma_5 
\sum_{f} N_c^{(f)} \delta_p^{(f)} \Bigg\} \non \\
&\equiv& i 
g^1_{Z \tilde \chi_1^0 \tilde \chi_1^0} \gamma_\mu \gamma_5 + i
g^p_{Z \tilde \chi_1^0 \tilde \chi_1^0} (p_1-p_2)_\mu \gamma_5 ,
\eeq
with the color factors $N_c^{(f)}$, and
\beq \label{zloop}
\delta_a^{(f)} &=&
(a_f+v_f c_{2\theta_{\tilde{f}}}) v_1 \left[ 
- m_f^2 C_0(\sqa) -  \mx^2 C_0(\sqb) + 4 \mx^2 C_1^+(\sqb) \right. \non
\\
& &- \left. 
2(\mx^2+p_1 \cdot p_2) C_2^+(\sqb) -2 (\mx^2-p_1\cdot p_2) C_2^-(\sqb)
-2 C_2^0(\sqb) \right] \non \\
&+ & (a_f-v_f c_{2\theta_{\tilde{f}}}) v_1 \left[ 
- m_f^2 C_0(\sqb) -  \mx^2 C_0(\sqa) + 4 \mx^2 C_1^+(\sqa) \right.
\non \\
& & - \left. 
2(\mx^2+p_1 \cdot p_2) C_2^+(\sqa) -2 (\mx^2-p_1\cdot p_2) C_2^-(\sqa)
-2 C_2^0(\sqa) \right] \non \\
&+& 2 m_f \mx a_f \sq (v_1+v_3) \left[ C_0(\sqa) - C_0(\sqb) -2C_1^+(\sqa) 
+ 2C_1^+(\sqb)\right] \non \\
&- & (a_f+v_f) v_2 m_f^2 \left[ C_0(\sqa) c^2_{\theta_{\tilde{f}}} +
C_0(\sqb) s^2_{\theta_{\tilde{f}}} \right] \non\\
& +& (a_f-v_f) v_2 \left[ c^2_{\theta_{\tilde{f}}} \left( -\mx^2 C_0(\sqa) 
+4\mx^2 C_1^+(\sqa) -2 C_2^0(\sqa) \right. \right. \non \\
& & \ \ \ - \left. 2(\mx^2+p_1\cdot p_2) C_2^+(\sqa) 
-2(\mx^2-p_1\cdot p_2) C_2^-(\sqa) \right) \non\\
& &+ \ s^2_{\theta_{\tilde{f}}} \left( -\mx^2 C_0(\sqb) 
+4\mx^2 C_1^+(\sqb) -2 C_2^0(\sqb) \right. \non \\
&& \ \ \ - \left. \left. 2(\mx^2+p_1\cdot p_2) C_2^+(\sqb) 
-2(\mx^2-p_1\cdot p_2) C_2^-(\sqb) \right) \right] \non \\
&-& \cq v_1 \left[ f_{Z\sqa\sqa} C_2^0(\sqa,\sqa) - f_{Z\sqb\sqb} 
C_2^0(\sqb,\sqb) \right] \non\\ 
&-& v_2 \left[ c^2_{\theta_{\tilde{f}}}f_{Z\sqa\sqa} C_2^0(\sqa,\sqa) 
+ s^2_{\theta_{\tilde{f}}} f_{Z\sqb\sqb} C_2^0(\sqb,\sqb) \right] \non\\ 
&+& \sq (2v_1+v_2) f_{Z\sqa\sqb} C_2^0(\sqa,\sqb) ;
\eeq
\beq
\delta_p^{(f)} &=& (a_f+v_f\cq) 2\mx v_1 \left[ C_1^+(\sqb) 
- 2 C_2^-(\sqb) \right]
\non \\
&+& (a_f-v_f\cq) 2\mx v_1 \left[ C_1^+(\sqa) - 2 C_2^-(\sqa) \right] \non\\
&+& 2 a_f\sq (v_1+v_3) m_f\left[ -C_1^+(\sqa) + C_1^+(\sqb)\right] \non \\
&+& 2(a_f-v_f) v_2 \mx\left[ c^2_{\theta_{\tilde{f}}} \left( C_1^+(\sqa) - 2
C_2^-(\sqa)\right) 
+ s^2_{\theta_{\tilde{f}}} \left( C_1^+(\sqb) - 2 C_2^-(\sqb) 
\right) \right] \non\\ 
&+& 2\cq v_1 \mx\left[ f_{Z\sqa\sqa} C_2^-(\sqa,\sqa) - f_{Z\sqb\sqb} 
C_2^-(\sqb,\sqb) \right] \non\\
&+& 2 v_2\mx \left[ c^2_{\theta_{\tilde{f}}} f_{Z\sqa\sqa} C_2^-(\sqa,\sqa) 
+ s^2_{\theta_{\tilde{f}}} f_{Z\sqb\sqb}  C_2^-(\sqb,\sqb) \right] \non\\
&+& 2 f_{Z\sqa\sqb} \left[ (v_1+v_3) m_f C_1^-(\sqa,\sqb) -
\sq (2v_1 + v_2)\mx C_2^-(\sqa,\sqb) \right].
\eeq
Here we have used the notation 
\beq \label{gauginocoup}
v_f &=& -\frac{1}{2} I_3^f + s^2_W Q_f \ \  , \ \ 
a_f \ = \ \frac{1}{2} I_3^f, \hspace*{5cm} \non \\
v_0 &=& \left( g Q_f N_{11} \tan\theta_W \right)^2 \ , \ 
v_1 \ = \ \frac{1}{2} v_0, \non \\
v_2 &=& \frac{I_3^f}{2 Q_f} v_0 \left[ -2 +\frac{I_3^f}{Q_f}
 + 2 \left( 1 - \frac{I_3^f}{Q_f} \right) \frac{N_{12}} {N_{11} \tan \theta_W}
 + \frac{I_3^f}{Q_f} \left( \frac{N_{12}} {N_{11} \tan \theta_W} \right)^2 
 \right] , \non \\
v_3 &=&  \frac{I_3^f}{2 Q_f} v_0 \left( \frac {N_{12}} {N_{11} 
\tan \theta_W} -1 \right).
\eeq
The $Z$ boson couplings to sfermions are given by: 
\beq \label{zcoup}
f_{Z\sqa\sqa} = -2I_{3}^f c^2_{\theta_{\tilde{f}}}+2 Q_f s^2_W \ , \
f_{Z\sqb\sqb} = -2I_{3}^f s^2_{\theta_{\tilde{f}}}+2 Q_f s^2_W \ , \ 
f_{Z\sqa\sqb} = I_{3}^f \sq . 
\eeq
The angle $\theta_{\tilde f}$ is the mixing angle arising from the
diagonalization of the sfermion mass matrices [compare
eq.~(\ref{sqstate}) below] and $s^2_{\theta_{\tilde f}} =
1-c^2_{\theta_{\tilde f}} \equiv \sin^2\theta_{\tilde f}$. $Q_f$ and
$I_3^f$ denote the electric charge and weak isospin of the fermion
$f$, respectively. The Passarino--Veltman three--point functions
\cite{R6}, defined as
\beq \label{loopfun}
C_{0,1,2}^{0,+,-} (\tilde{f}) &\equiv & C_{0,1,2}^{0,+,-} 
(q^2, m_{\tilde\chi_1^0}^2, m_f^2, m_f^2, m_{\tilde{f}}^2) ; \non \\
C_{0,1,2}^{0,+,-} (\tilde{f}_1, \tilde{f}_2) & \equiv & C_{0,1,2}^{0,+,-} 
(q^2, m_{\tilde\chi_1^0}^2, m_{\tilde{f}_1}^2, m_{\tilde{f}_2}^2, m_f^2),
\eeq
with $q$ the momentum of the Higgs or $Z$ boson, can be found in
Ref.~\cite{higgsino}.\footnote{Some care has to be taken if $q^2
\rightarrow 0$ or $q^2 = 4 m_{\tilde \chi_1^0}^2$, since standard
expressions for the loop functions contain spurious divergences in
these kinematical situations, which are characteristic for
LSP--nucleus scattering and LSP annihilation, respectively. This
problem is discussed in the Appendix of Ref.~\cite{higgsino}.}\s

In the same notation, the one--loop Higgs boson couplings to the LSP 
neutralinos in the bino limit are given by:
\beq \label{goneh}
\Gamma^1(\phi\tilde{\chi}_1^0\tilde{\chi}_1^0) &=& \frac{ig}{4\pi^2} \bigg[  
\sum_f N_c \delta_\phi^{(f)} \bigg] \equiv
i g^1_{\phi \tilde \chi_1^0 \tilde \chi_1^0}
\non \\
\Gamma^1(A\tilde{\chi}_1^0\tilde{\chi}_1^0) &=& 
\frac{g}{4\pi^2} \gamma_5 \bigg[ \sum_f N_c \delta_A^{(f)} \bigg] 
\equiv g^1_{A \tilde \chi_1^0 \tilde \chi_1^0} \gamma_5
\eeq
where
\beq \label{hloop}
\delta_\phi^{(f)} &=& \frac{m_f g_{\phi ff}}{2m_W} \left\{ 
\sq (v_1+v_3)\left[ - \left(\msqa^2 + m_f^2 + \mx^2 \right) C_0(\sqa) +
4 \mx^2 C_1^+(\sqa) \right.\right. \non\\
&& \ \ \ + \left. \left(\msqb^2 + m_f^2 + \mx^2 \right) C_0(\sqb) - 
4 \mx^2 C_1^+(\sqb) \right] \non\\
&& + \ 2(v_1+v_2 c^2_{\theta_{\tilde{f}}}) m_f\mx 
\left[ C_0(\sqa) - 2C_1^+(\sqa) \right] \non \\
&& + \left. 2(v_1+v_2 s^2_{\theta_{\tilde{f}}}) m_f\mx \left[ C_0(\sqb) 
- 2C_1^+(\sqb) \right] \right\} \non \\
&-& C_{\phi\sqa\sqa} \left\{ -\sq (v_1+v_3) m_f C_0(\sqa,\sqa) + 
2(v_1+v_2 c^2_{\theta_{\tilde{f}}}) \mx C_1^+(\sqa,\sqa) \right\} \non\\
&-& C_{\phi\sqb\sqb} \left\{ \sq (v_1+v_3) m_f C_0(\sqb,\sqb) + 
2(v_1+v_2 s^2_{\theta_{\tilde{f}}}) \mx C_1^+(\sqb,\sqb) \right\} \non\\
&-& C_{\phi\sqa\sqb} \left\{ -2 \cq (v_1+v_3) m_f C_0(\sqa,\sqb) 
 -2 \sq v_2\mx C_1^+(\sqa,\sqb) 
 \right\};
\eeq
\beq
\delta_A^{(f)} &=& \frac{m_f g_{Aff}}{2 m_W} \left\{ (v_1+v_3) \sq 
\left[\left( \msqa^2 - m_f^2 -\mx^2 \right) C_0(\sqa) - 
\left( \msqb^2 - m_f^2 -\mx^2 \right) C_0(\sqb) \right] \right. \non \\
&& + \left. 2(v_1+v_2 c^2_{\theta_{\tilde{f}}}) \mx m_f C_0(\sqa) + 
2(v_1+v_2 s^2_{\theta_{\tilde{f}}})\mx m_f C_0(\sqb) \right\} \non\\
&+& C_{A\sqa\sqb} \left\{ -2 m_f (v_1+v_3) C_0(\sqa,\sqb) 
+ 2\mx (2v_1 + v_2) \sq C_1^-(\sqa,\sqb) \right\} .
\eeq
The Higgs--fermion--fermion coupling constants are given by
\beq \label{hffcoup}
g_{huu} &=& \frac{\cos\alpha}{\sin\beta} \qquad , \qquad
g_{hdd} = \frac{-\sin\alpha}{\cos\beta}\ , \non\\
g_{Huu} &=& \frac{\sin\alpha}{\sin\beta} \qquad , \qquad
g_{Hdd} = \frac{\cos\alpha}{\cos\beta} \ , \non\\
g_{Auu} &=& \cot\beta \qquad , \qquad g_{Add} = \tan\beta \ ,
\eeq
while the Higgs--sfermion--sfermion coupling constants read:
\beq \label{hsfsfcoup}
C_{h\tilde{u}_1\tilde{u}_1} &=& \frac{m_Z}{c_W}
s_{\beta+\alpha} \left[ I_{3}^u c^2_{\theta_{\tilde{u}}}  
-Q_u s^2_W c_{2\theta_{\tilde{u}}}\right] 
-\frac{m_u^2 g_{huu}}{m_W} -
\frac{m_u s_{2\theta_{\tilde{u}}}}{2m_W} 
\left[ A_u g_{huu} + \mu g_{Huu} \right]; \non\\
C_{h\tilde{u}_2\tilde{u}_2} &=& \frac{m_Z}{c_W}
s_{\beta+\alpha} \left[ I_{3}^u s^2_{\theta_{\tilde{u}}}
+ Q_u s^2_W c_{2\theta_{\tilde{u}}}\right] 
-\frac{m_u^2 g_{huu}}{m_W} +
\frac{m_u s_{2\theta_{\tilde{u}}}}{2m_W} \left[ A_u g_{huu} +
\mu g_{Huu} \right]; \non\\
C_{h\tilde{u}_1\tilde{u}_2} &=& \frac{m_Z}{c_W}
s_{\beta+\alpha} \left[ Q_u s^2_W
-I_{3}^u/2 \right] s_{2\theta_{\tilde{u}}}
-\frac{m_u}{2m_W} \left[ A_u g_{huu} +
\mu g_{Huu} \right] c_{2\theta_{\tilde{u}}}; \non\\
%
%
C_{h\tilde{d}_1\tilde{d}_1} &=& \frac{m_Z}{c_W}
s_{\beta+\alpha} \left[ I_{3}^d c^2_{\theta_{\tilde{d}}}
-Q_d s^2_W c_{2\theta_{\tilde{d}}}\right] 
-\frac{m_d^2 g_{hdd}}{m_W} -
\frac{m_d s_{2\theta_{\tilde{d}}}}{2m_W} 
\left[ A_d g_{hdd} - \mu g_{Hdd} \right]; \non\\
C_{h\tilde{d}_2\tilde{d}_2} &=& \frac{m_Z}{c_W}
s_{\beta+\alpha} \left[ I_{3}^d s^2_{\theta_{\tilde{d}}}
+ Q_d s^2_W c_{2\theta_{\tilde{d}}}\right] 
-\frac{m_d^2 g_{hdd}}{m_W} +
\frac{m_d s_{2\theta_{\tilde{d}}}}{2m_W} \left[ A_d g_{hdd} -
\mu g_{Hdd} \right]; \non\\
C_{h\tilde{d}_1\tilde{d}_2} &=& \frac{m_Z}{c_W}
s_{\beta+\alpha} \left[ Q_d s^2_W-I_{3}^d/2 \right] s_{2\theta_{\tilde{d}}} 
-\frac{m_d}{2m_W} \left[ A_d g_{hdd} -
\mu g_{Hdd} \right] c_{2\theta_{\tilde{d}}}; \non\\
%
%
C_{H\tilde{u}_1\tilde{u}_1} &=& -\frac{m_Z}{c_W}
c_{\beta+\alpha} \left[ I_{3}^u c^2_{\theta_{\tilde{u}}}
-Q_u s^2_W c_{2\theta_{\tilde{u}}}\right] 
-\frac{m_u^2 g_{Huu}}{m_W} -
\frac{m_u s_{2\theta_{\tilde{u}}}}{2m_W} \left[ A_u g_{Huu} -
\mu g_{huu} \right]; \non\\
C_{H\tilde{u}_2\tilde{u}_2} &=& -\frac{m_Z}{c_W}
c_{\beta+\alpha} \left[ I_{3}^u s^2_{\theta_{\tilde{u}}}
+Q_u s^2_W c_{2\theta_{\tilde{u}}}\right] 
-\frac{m_u^2 g_{Huu}}{m_W} +
\frac{m_u s_{2\theta_{\tilde{u}}}}{2m_W} \left[ A_u g_{Huu} -
\mu g_{huu} \right]; \non\\
C_{H\tilde{u}_1\tilde{u}_2} &=& -\frac{m_Z}{c_W}
c_{\beta+\alpha} \left[ Q_u s^2_W-I_{3}^u/2 \right] s_{2\theta_{\tilde{u}}} 
-\frac{m_u}{2m_W} \left[ A_u g_{Huu} -
\mu g_{huu} \right] c_{2\theta_{\tilde{u}}}; \non\\
%
%
C_{H\tilde{d}_1\tilde{d}_1} &=& -\frac{m_Z}{c_W}
c_{\beta+\alpha} \left[ I_{3}^d c^2_{\theta_{\tilde{d}}}
-Q_d s^2_W c_{2\theta_{\tilde{d}}}\right] 
-\frac{m_d^2 g_{Hdd}}{m_W} -
\frac{m_d s_{2\theta_{\tilde{d}}}}{2m_W} \left[ A_d g_{Hdd} +
\mu g_{hdd} \right]; \non\\
C_{H\tilde{d}_2\tilde{d}_2} &=& -\frac{m_Z}{c_W}
c_{\beta+\alpha} \left[ I_{3}^d s^2_{\theta_{\tilde{d}}}
+Q_d s^2_W c_{2\theta_{\tilde{d}}}\right] 
-\frac{m_d^2 g_{Hdd}}{m_W} +
\frac{m_d s_{2\theta_{\tilde{d}}}}{2m_W} \left[ A_d g_{Hdd} +
\mu g_{hdd} \right]; \non\\
C_{H\tilde{d}_1\tilde{d}_2} &=& -\frac{m_Z}{c_W}
c_{\beta+\alpha} \left[ Q_d s^2_W-I_{3}^d/2 \right] s_{2\theta_{\tilde{d}}} 
-\frac{m_d}{2m_W} \left[ A_d g_{Hdd} +
\mu g_{hdd} \right] c_{2\theta_{\tilde{d}}}; \non\\
%
%
C_{A\tilde{u}_1\tilde{u}_2} &=& \frac{m_u}{2m_W} (A_u \cot\beta +\mu);
 \non\\
C_{A\tilde{d}_1\tilde{d}_2} &=& \frac{m_d}{2m_W} (A_d \tan\beta +\mu).
\eeq

We use the following convention for the sfermion mass matrices:
\beq \label{sqmass_matrix}
{\cal M}^2_{\tilde{f}} =
\left(
  \begin{array}{cc} m_f^2 + m_{LL}^2 & m_f \, \tilde{A}_f  \\
                    m_f\, \tilde{A}_f    & m_f^2 + m_{RR}^2 
  \end{array} \right), \ \ {\rm with} 
\begin{array}{l} 
\ m_{LL}^2 =m_{\tilde{f}_L}^2 + (I_3^f - Q_f s_W^2)\, m_Z^2\, c_{2\beta} \\\
m_{RR}^2 = m_{\tilde{f}_R}^2 + Q_f s_W^2\, m_Z^2\, c_{2\beta} \\\
\ \tilde{A}_f  = A_f - \mu (\tb)^{-2 I_3^f} 
\end{array} .
\eeq
They are diagonalized by $ 2 \times 2$ rotation matrices described by
the angles $\theta_{\tilde{f}}$, which turn the current eigenstates,
$\tilde{f}_L$ and $\tilde{f}_R$, into the mass eigenstates
$\tilde{f}_1$ and $\tilde{f}_2$; the mixing angle and sfermion masses
are then given by
\beq \label{sqstate}
s_{2\theta_{\tilde{f}}} = \frac{2 m_f \tilde{A}_f} { m_{\tilde{f}_1}^2
-m_{\tilde{f}_2}^2 } \ \ , \ \ 
c_{2\theta_{\tilde{f}}} = \frac{m_{LL}^2 -m_{RR}^2} 
{m_{\tilde{f}_1}^2 -m_{\tilde{f}_2}^2 }, \hspace*{0.8cm}  \\
m_{\tilde{f}_{1,2}}^2 = m_f^2 +\frac{1}{2} \left[
m_{LL}^2 +m_{RR}^2 \mp \sqrt{ (m_{LL}^2
-m_{RR}^2 )^2 +4 m_f^2 \tilde{A}_f^2 } \ \right] .
\eeq

\section{Higgs and Z Boson Decays}

\subsection{Invisible decays of the Z boson}

The tree level and the one--loop induced axial vector couplings of the
$Z$ boson to \lsp\ pairs are shown in Fig.~2a for the input values
$\tb=15$ and $\mu=1$ TeV.  We work in an unconstrained model with
non--universal boundary conditions for the gaugino mass parameters at
the high energy scale,\footnote{In models with universal gaugino
masses at the GUT scale, the bound on the lightest chargino mass from
negative searches at LEP2, $m_{\tilde{\chi}_1^\pm} \gsim 104$ GeV
\cite{R7,R8}, will constrain a bino--like neutralino to have a mass
larger than $m_Z/2$; the decay $Z \ra \tilde{\chi}_1^0
\tilde{\chi}_1^0$ would then be kinematically closed.} and set the
bino and wino mass parameters to $M_1=30$ GeV and $M_2=120$ GeV,
respectively. For definiteness we assume a common soft SUSY breaking
scalar mass for the three generations of sleptons, $m_{\tilde{e}_L} =
m_{\tilde{e}_R} \equiv m_{\tilde l}$, and a common mass term for the
squarks $m_{\tilde{q}_L} = m_{\tilde{u}_R} = m_{\tilde{d}_R} \equiv
m_{\tilde q}$, with $m_{\tilde q}=2m_{\tilde l}$. For the trilinear
coupling, which will play a role mainly in the stop sector, we choose
the value $A_t = 2.9 m_{\tilde{q}}$. The trilinear couplings in the
$\tilde d-$squark and slepton sectors, which are not important here,
have been set to zero. \s

\begin{figure}[htb]
\begin{center}
\epsfig{figure=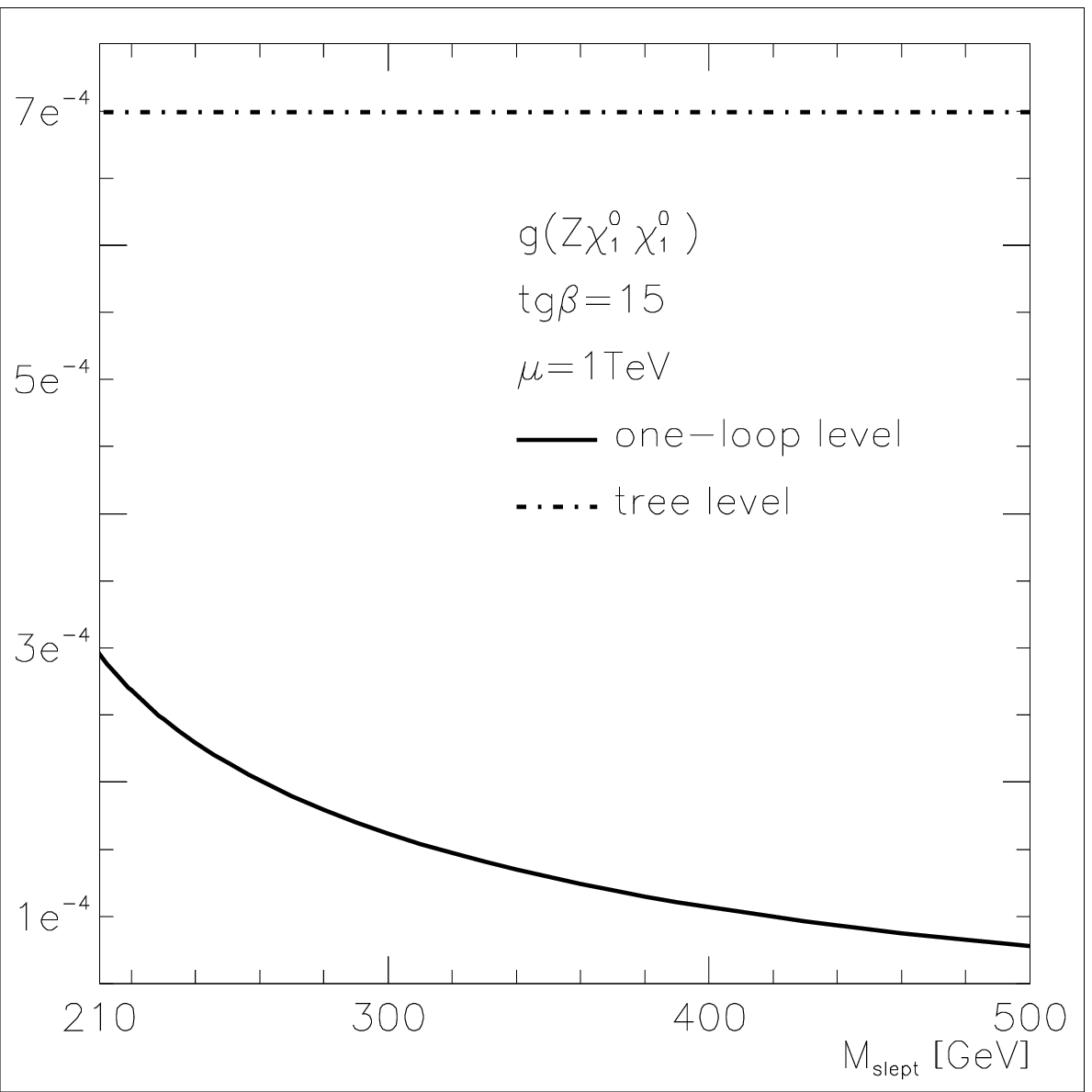,bbllx=2pt,bblly=2,bburx=338,bbury=338, 
width=7.5cm,clip=}
\hspace{1.cm}
\epsfig{figure=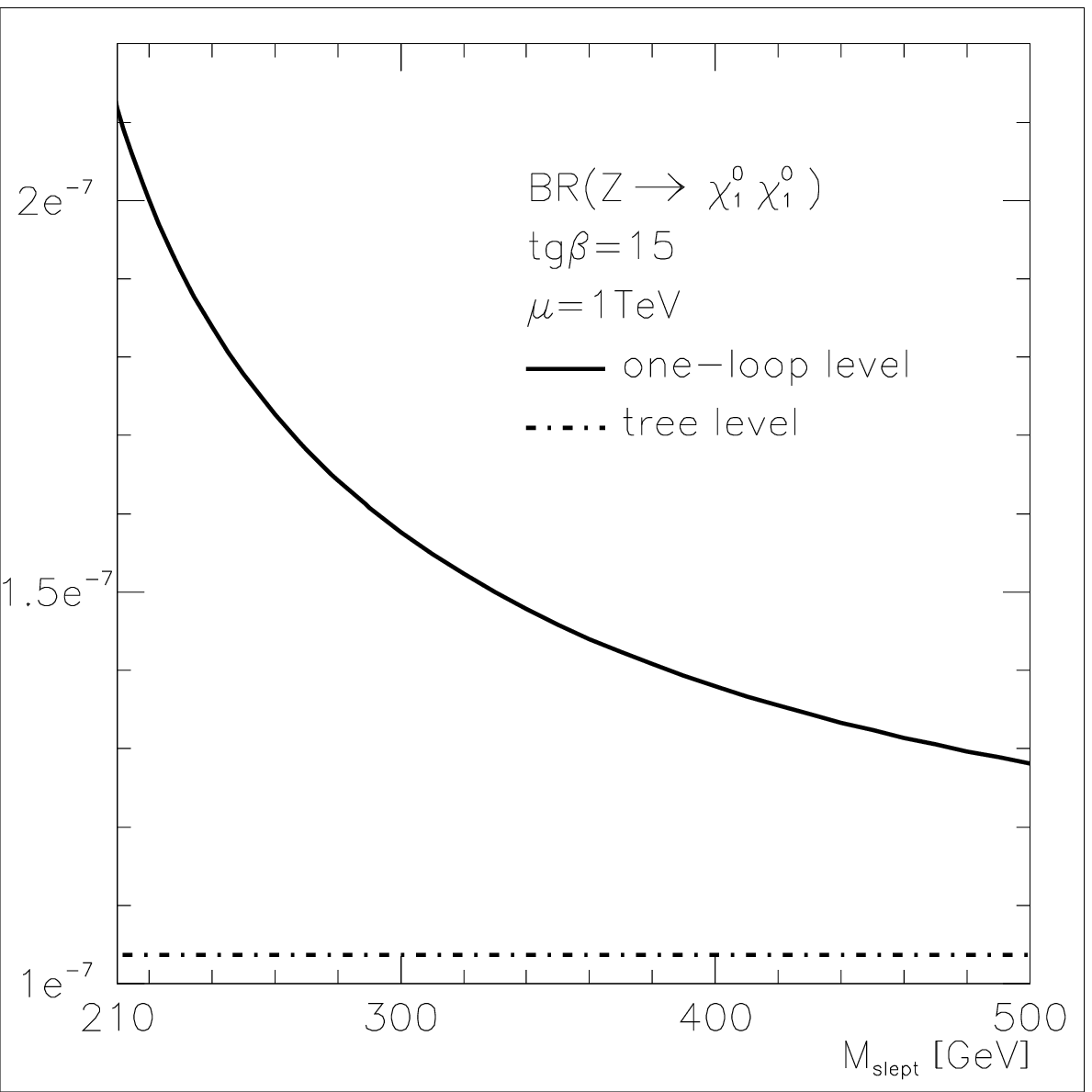,bbllx=2pt,bblly=2,bburx=338,bbury=338,
width=7.5cm,clip=}
\end{center}
\vspace*{-4mm}
\caption{The $Z$ boson axial vector coupling (left) and branching
ratio (right) to pairs of the lightest neutralino as functions of the
common slepton mass. The dashed lines show the tree--level results; the
solid line in the left frame shows the real part of the one--loop
contribution eq.~(\ref{gonez}) to the axial vector coupling, while the
solid line in the right frame shows the total loop--corrected
branching ratio eq.~(\ref{zdec}). These results are given for $\tb=15, \mu
=1$ TeV, $m_{\tilde{q}} = 2 m_{\tilde{l}}, A_t = 2.9 m_{\tilde{q}}$
and gaugino masses $M_1=30$ GeV, $M_2=120$ GeV.}
\end{figure}

The tree--level coupling $g^0_{ Z \tilde{\chi}_1^0 \tilde{\chi}_1^0}$
is very small if \lsp\ is bino--like, being of ${\cal O}
(m_Z^2/\mu^2)$. As can be seen from Fig.~2a, it is less than $10^{-3}$
for the chosen parameters [and of course does not depend on slepton
and squark masses]. The real part of the one--loop induced coupling,
$g^1_{ Z \tilde{\chi}_1^0 \tilde{\chi}_1^0}$, for on--shell $Z$ bosons
[i.e. with $q^2 =m_Z^2$] receives contributions from both massless and
massive SM fermions and their superpartners. In the former case the
real part has an ``accidental'' zero for sfermion masses slightly
above the $Z$ mass; this happens e.g. for the (s)neutrino contribution
at $m_{\tilde \nu} \simeq 94$ GeV.\footnote{This statement holds,
strictly speaking, only in the $\overline{\rm MS}$ scheme, where the
contribution from each (s)fermion species decouples separately when
$m_{\tilde f} \rightarrow \infty$. In the $\overline{\rm DR}$ scheme
decoupling only occurs after summation over a complete generation.}
The real part of this contribution is positive (negative) for smaller
(larger) sneutrino masses; its absolute value reaches a maximum at
$m_{\tilde \nu} \simeq 155$ GeV. No such zero occurs for the imaginary
parts. Moreover, the (s)top loop contribution is much larger than that
from the other (s)quarks, even if the stop masses are not reduced
significantly compared to the other squark masses. When combined with
the ``accidental'' suppression of the (s)lepton contribution for
slepton masses close to their present lower experimental bound, this
implies that the dominant contribution to the real part of this
coupling usually comes from (s)top loops. Of course, these loops
cannot contribute to the imaginary part, which receives its dominant
contribution from (s)lepton loops. For large masses, sparticles
decouple and the loop correction to the coupling vanishes
asymptotically. \s

The partial decay width for the decay of a $Z$ boson into a pair of
LSP neutralinos is given by:
\beq \label{zdec}
\Gamma (Z \ra \tilde{\chi}_1^0\tilde{\chi}_1^0) &=& \frac{\beta_Z^3 m_Z}{24\pi}
\left| g^0_{Z\tilde{\chi}_1^0\tilde{\chi}_1^0} + g^1_{Z\tilde{\chi}_1^0
\tilde{\chi}_1^0} \right|^2,
\eeq
where $\beta_Z=(1- 4 m_{\tilde{\chi}_1^0}^2/m_Z^2)^{1/2}$ is the
velocity of the neutralinos in the rest frame of the $Z$ boson. In the
usual perturbative expansion the one--loop correction to a decay width
originates from the interference of tree--level diagrams with the real
parts of the corresponding one--loop diagrams. However, in our case
the tree--level decay width will vanish in the limit $|\mu|
\rightarrow \infty$. In this limit the one--loop diagrams represent
the leading order contributions. We therefore also include the squared
one--loop contribution in eq.~(\ref{zdec}). However, for $|\mu| \lsim
1$ TeV the numerically most important correction to the tree--level
result usually still comes from the interference between the tree
level and the real part of the one--loop couplings. Note finally that
the derivative coupling $g^p_{Z \tilde\chi_1^0 \tilde\chi_1^0}$ of
eq.~(\ref{gonez}) does not contribute to the decay width of real $Z$
bosons, since the product of the $Z$ polarization vector with the sum
of the outgoing LSP momenta vanishes. \s

The $Z$ decay branching ratio, BR$(Z \ra \tilde{\chi}_1^0
\tilde{\chi}_1^0) =\Gamma (Z \ra \tilde{\chi}_1^0 \tilde{\chi}_1^0) / 
\Gamma_Z^{\rm tot}$ with $\Gamma_Z^{\rm tot} = 2.5$ GeV, is shown in 
Fig.~2b. In the Born approximation, the branching ratio is very small,
$\sim 1 \cdot 10^{-7}$. It can be enhanced by the loop contribution
by a factor of $\gsim 2$ in the low slepton mass range. However, the
branching fraction is still well below the experimental limit on the
invisible $Z$ boson decay width \cite{R7},
\beq \label{zexp}
\Delta \Gamma_Z^{\rm inv}|_{\rm exp}  \simeq \pm 1.5 \ {\rm MeV} ,
\eeq
which would require a branching ratio in the permille range for
detection. One can find scenarios with somewhat larger loop
corrections than shown in Fig.~2, if parameters are chosen such that
all slepton masses are near 150 GeV. However, even in this case we are
still more than three orders of magnitude below the experimental limit
(\ref{zexp}). Of course the $Z$ partial decay width into \lsp\ pairs
can be enhanced to a detectable level for lower values of the
parameter $\mu$. However, in this case the one--loop corrections
[which remain more or less the same] would be relatively less
important than for large $|\mu|$.

\subsection{Invisible decays of the Higgs bosons} 

\begin{figure}
\begin{center}
\epsfig{figure=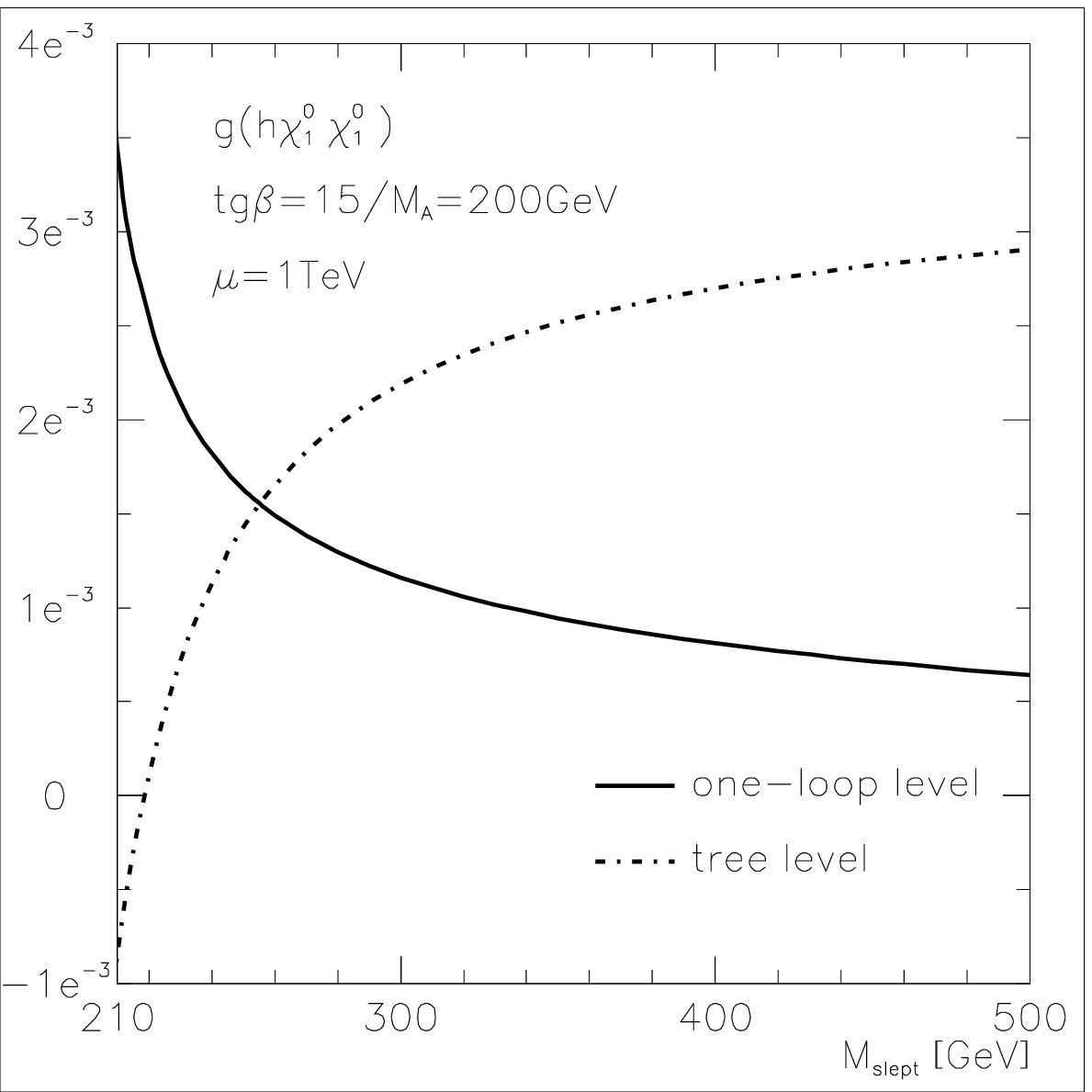,bbllx=2pt,bblly=2,bburx=338,bbury=338,
width=7.5cm,clip=}
\hspace{1.cm}
\epsfig{figure=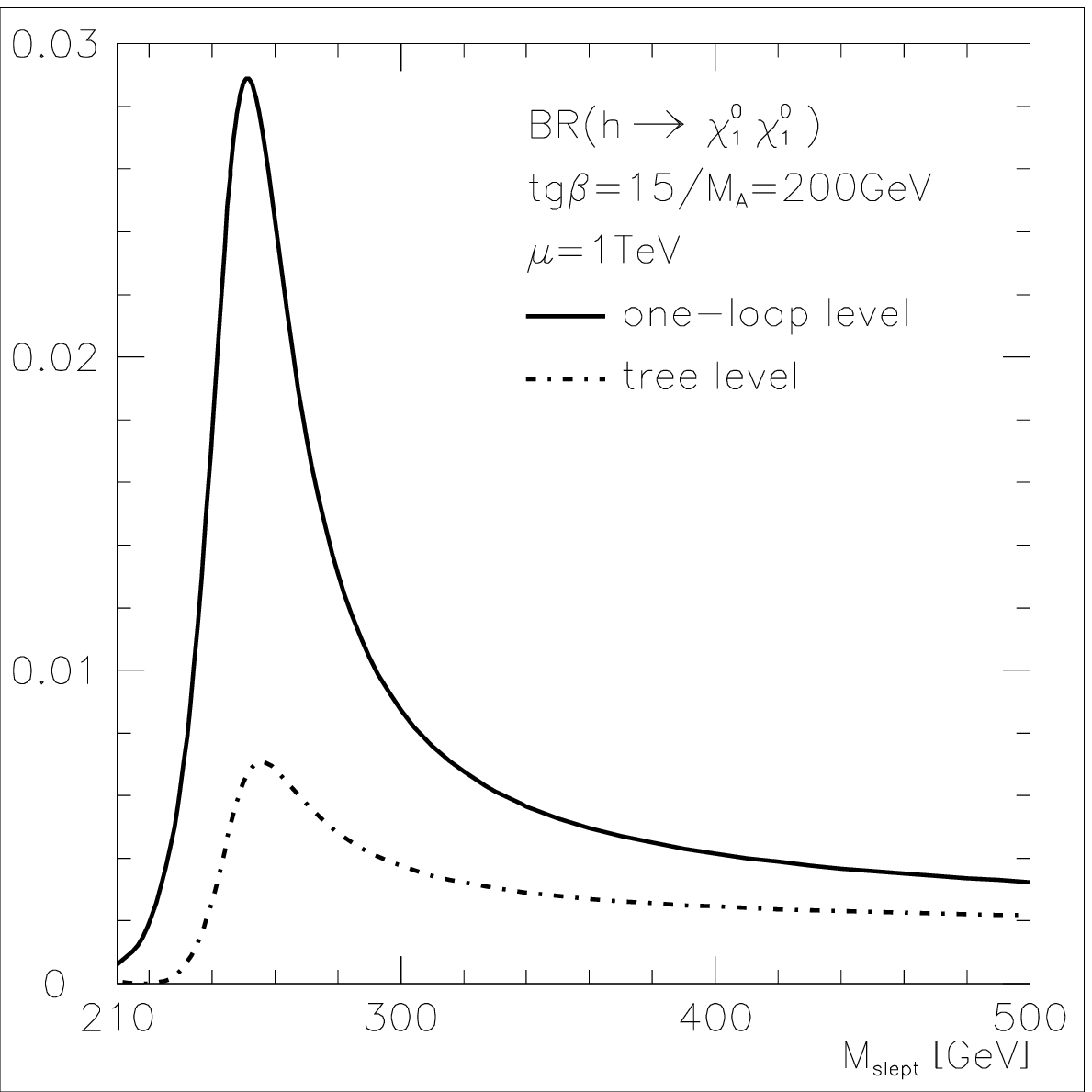,bbllx=2pt,bblly=2,bburx=338,bbury=338,
width=7.5cm,clip=}\\[0.2cm]
\end{center}
\vspace*{1cm}
\begin{center}
\epsfig{figure=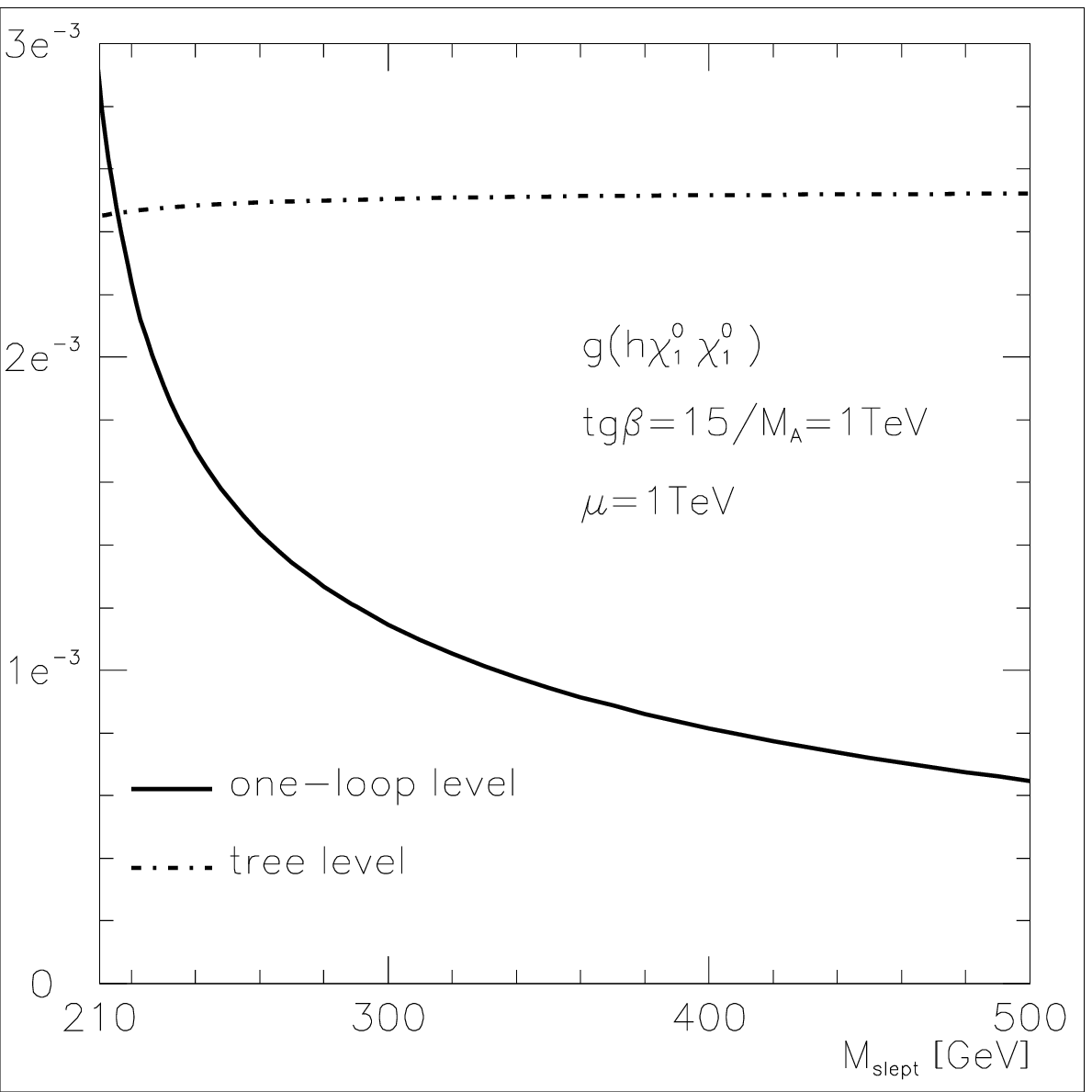,bbllx=2pt,bblly=2,bburx=338,bbury=338,
width=7.5cm,clip=}
\hspace{1.cm}
\epsfig{figure=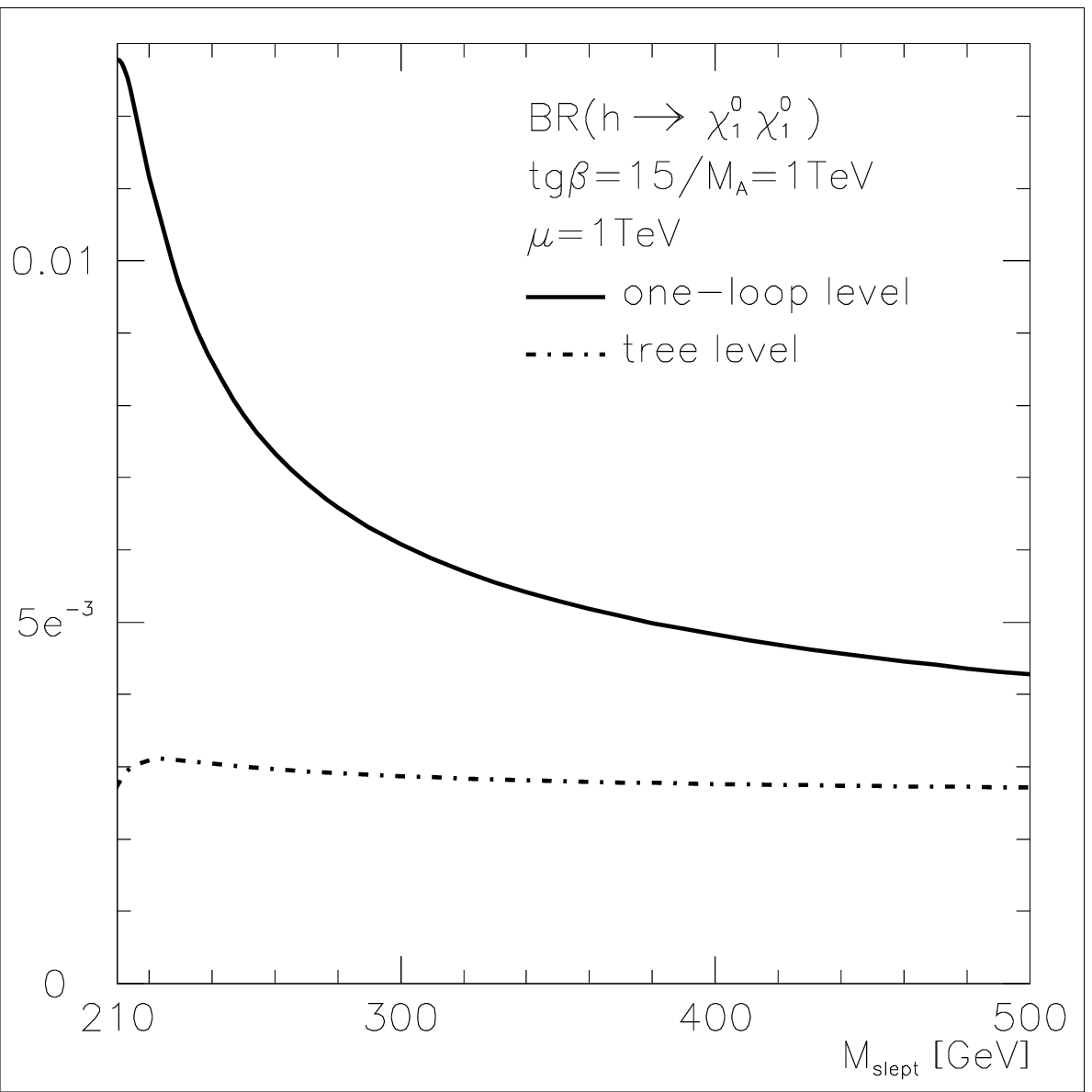,bbllx=2pt,bblly=2,bburx=338,bbury=338,
width=7.5cm,clip=}
\end{center}
\vspace*{.5cm}
\caption{The lightest $h$ boson couplings (left) and branching ratios (right) 
to pairs of the lightest neutralinos as functions of the common slepton mass. 
Most parameters are as in Fig.~2, and we took $M_A=200$ GeV (top) and 1 TeV 
(bottom).}
\end{figure}

The real part of the coupling of an on--shell lightest $h$ boson to a
\lsp\ pair is displayed, at the Born and one--loop level, in the
left--hand frames of Figs.~3. It is again shown as a function of the
sfermion masses, for $\tb=15$, $\mu=1$ TeV and pseudoscalar mass input
values $M_A=200$ GeV (upper frames) and 1 TeV (lower frames). We see
that the tree--level coupling $g^0_{h \tilde{\chi}_1^0
\tilde{\chi}_1}$ is typically a factor of three larger than the
corresponding coupling $g^0_{Z \tilde{\chi}_1^0 \tilde{\chi}_1^0}$ of
the $Z$ boson. This can be understood from eqs.~(\ref{zcoup0}),
(\ref{hcoup0}) and (\ref{binostate}). In the ``decoupling'' limit
$M_A^2 \gg m_Z^2$ the neutral Higgs mixing angle $\alpha$ satisfies
$s_\alpha = -c_\beta, \ c_\alpha = s_\beta$. In this limit the ratio
of tree--level $Z \lsp \lsp$ and $h \lsp \lsp$ couplings becomes $-
m_Z \cot 2\beta / (2 \mu)$; for the given choice of parameters this
amounts to about 0.3, in good agreement with Figs.~2 and 3. On the
other hand, for small sfermion masses the one--loop corrections to the
$h \lsp \lsp$ coupling exceed $g^1_{Z \tilde{\chi}_1^0
\tilde{\chi}_1^0}$ by more than an order of magnitude. The corrections
to the $Z$ coupling to bino--like neutralinos only involve electroweak
gauge couplings. In contrast, top quarks couple with ${\cal O}(1)$
Yukawa coupling to the Higgs bosons, and for the given choice of large
$|A_t|$ the (dimensionful) $h \tilde t_1 \tilde t_1$ coupling
significantly exceeds the $\tilde t_1$ mass. Moreover, due to $\tilde
t_L - \tilde t_R$ mixing the lighter $\tilde t$ mass eigenstate is
often not only lighter than the other squarks, but also lighter than
the sleptons. As a result, the loop corrections are numerically even
{\em more} important in case of the Higgs coupling, even though the
tree--level contribution to this coupling is nominally only suppressed
by one power of $|\mu|$. If $|A_t|$ is large, as in the present
example, the loop corrections to the $h \lsp \lsp$ coupling can even
exceed the tree--level contribution. \s

The variation of the one--loop contribution to the coupling is again
mostly due to the natural decrease with increasing masses of the
sfermions running in the loop, which decouple when they are much
heavier than the $h$ boson. For $m_{\tilde q} \simeq 420$ GeV
[i.e. $m_{\tilde l} \simeq 210$ GeV], $m_{\tilde t_1}$ is near its
experimental lower bound of $\sim 100$ GeV, due to strong $\tilde t_L -
\tilde t_R$ mixing. This implies that $m_{\tilde t_1}$ will grow
faster than linearly with increasing $m_{\tilde q}$, which explains
the very rapid decrease of the loop corrections. However, there is
also a variation of the tree--level coupling for $M_A=200$ GeV which,
at first sight, is astonishing. It is caused by the variation of the
mixing angle $\alpha$ in the CP--even Higgs sector, and to a lesser
extent by the variation of $M_h$, due to the strong dependence of
crucial loop corrections in the CP--even Higgs sector on the stop
masses.\footnote{We remind the reader that LEP Higgs searches would
completely exclude the MSSM in the absence of these corrections.} In
fact, for the set of input parameters with $M_A=200$ GeV at small
slepton masses, we are in the regime where $\sin \alpha$, which
appears in the $h \tilde{\chi}_1^0 \tilde{\chi}_1^0$ coupling [and
which enters the $h b\bar{b}$ coupling as will be discussed later],
varies very quickly. This ``pathological" region, where the
phenomenology of the MSSM Higgs bosons is drastically affected, has
been discussed in several places in the literature \cite{R9}.
\smallskip

The partial widths for the decays of CP--even Higgs bosons, $\phi=h,H$, into 
pairs of identical neutralinos are given by \cite{invisible}: 
\beq \label{hdec}
\Gamma (\phi \rightarrow \tilde{\chi}_1^0 \tilde{\chi}_1^0) &=&
\frac{\beta_\phi^3 M_\phi}{16\pi} 
\left| g^0_{\phi\tilde{\chi}_1^0\tilde{\chi}_1^0} +
 g^1_{\phi \tilde{\chi}_1^0 \tilde{\chi}_1^0} \right|^2.
\eeq
The branching ratios for the decays of the lightest $h$ boson are
shown in the right--hand frames of Figs.~3 for the same choice of
parameters previously discussed. They have been calculated by
implementing the one--loop Higgs couplings to neutralinos in the
Fortran code {\tt HDECAY} \cite{R10} which calculates all possible
decays of the MSSM Higgs bosons and where all important corrections in
the Higgs sector, in both the spectrum and the various decay widths,
are included. We see that the situation is completely different from
the case of the decay $Z \ra \tilde{\chi}_1^0\tilde{\chi}_1^0$: the
branching ratio BR$(h \ra \tilde{\chi}_1^0\tilde{\chi}_1^0)$ can
already exceed the one permille level with tree--level couplings. This
enhancement is not only due the larger $g_{h \tilde{\chi}_1^0
\tilde{\chi}_1^0}$ couplings compared to $g_{Z \tilde{\chi}_1^0
\tilde{\chi}_1^0}$, but also due to the much smaller $h$ total decay
width\footnote{For the choice of parameters in Fig.~3 the total decay
width of $h$ is similar to, or even smaller than, that of the Standard
Model Higgs boson with equal mass.}, $\Gamma_h^{\rm tot} \sim $ a few
MeV, compared to $\Gamma_Z^{\rm tot} \simeq 2.5$ GeV. After including
the one--loop corrections, the branching fraction BR$(h \ra
\tilde{\chi}_1^0 \tilde{\chi}_1^0)$ can be enhanced to reach the level
of a few percent. \smallskip

\begin{figure}
\begin{center}
\epsfig{figure=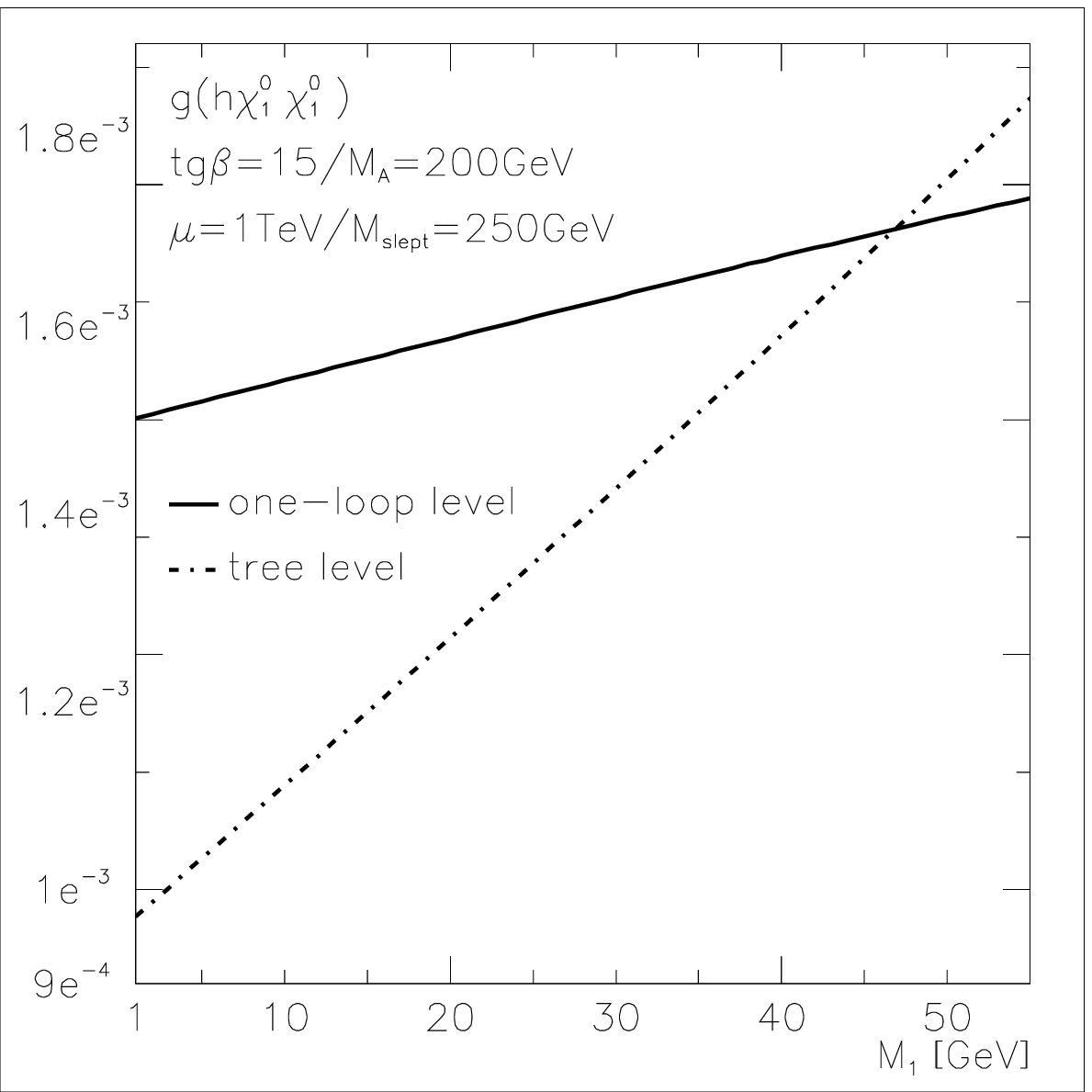,bbllx=2pt,bblly=2,bburx=338,bbury=338,
width=7.8cm,clip=}
\hspace{.5cm}
\epsfig{figure=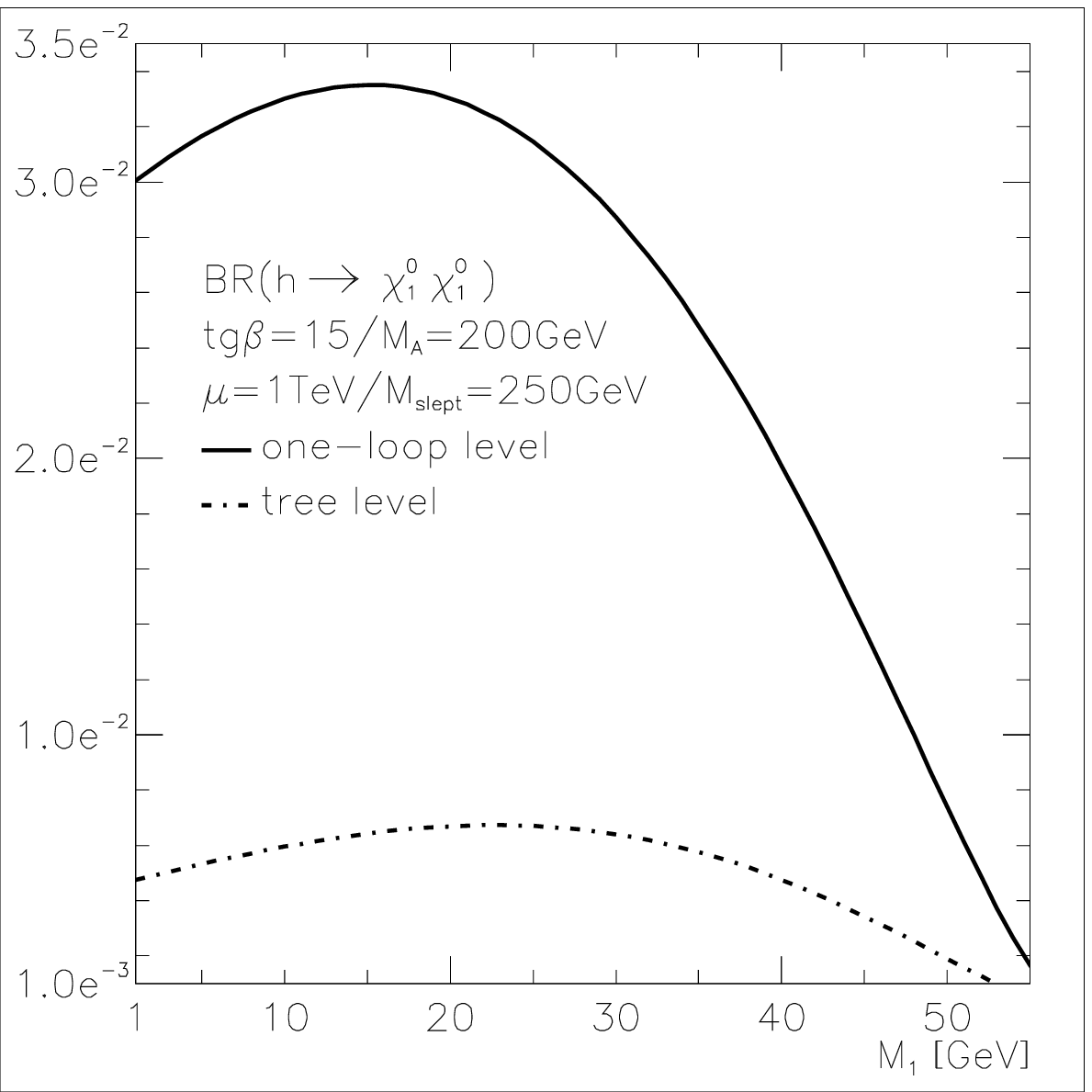,bbllx=2pt,bblly=2,bburx=338,bbury=338,
width=7.8cm,clip=}
\end{center}
\caption{The lightest $h$ boson couplings (left) and branching ratios (right) 
to pairs of the lightest neutralinos as functions of $M_1$. The parameters are 
as in Fig.~2 with $m_{\tilde{l}}=250$ GeV and $M_A = 200$ GeV.}
\end{figure}

The branching ratio is especially enhanced if the usually dominant
decay into $b \bar b$ pairs is suppressed, i.e. if $|\sin \alpha|$
is very small; recall that the $h b \bar b$ coupling is $\propto \sin
\alpha / \cos \beta$. In our examples this happens for $M_A=200$ GeV
and $m_{\tilde{l}} \simeq 250$ GeV. In this case $g^0_{h \tilde \chi_1^0
\tilde \chi_1^0}$ is only about half as large as in the decoupling
limit $M_A \rightarrow \infty$, but the loop contribution to this
coupling is still sizable for this value of the sfermion masses, and
has the same sign as the tree--level coupling, leading to a quite
large total coupling. The branching ratio falls off quickly for
smaller sfermion masses, since here $\sin \alpha$ becomes sizable (and
positive). Moreover, for $m_{\tilde l} \simeq 210$ GeV the tree--level
and one--loop contributions to the couplings have opposite sign. The
branching ratio also decreases when $m_{\tilde l}$ is raised above 250
GeV, albeit somewhat more slowly; here the rapid decrease of the loop
contribution is compensated by the increase of the tree--level
coupling, which however does not suffice to compensate the
simultaneous increase of $\sin^2 \alpha$. \s

Fig.~4 shows the dependence of the $h \lsp \lsp$ coupling and of the
corresponding $h$ branching ratio on the mass of the LSP, $m_{\tilde
\chi_1^0} \simeq M_1$. We see that the tree--level contribution to
this coupling depends essentially linearly on the LSP mass.
Eqs.~(\ref{hcoup0}) and (\ref{binostate}) show that, for the given
scenario where $c_\alpha \simeq s_\beta \simeq 1$, this linear
dependence on $M_1$ originates from $N_{14}$, where the contribution
with $M_1$ in the numerator is enhanced by a factor of $\tan\beta$
relative to the contribution with $\mu$ in the numerator. Therefore
the contribution $\propto M_1$ is not negligible even though in Fig.~4
we have $M_1 \ll |\mu|$. On the other hand, the one--loop contribution
to this coupling depends only very weakly on $M_1$. The small increase
of this contribution shown in Fig.~4 is mostly due to the explicit
$m_{\tilde \chi_1^0}$ dependence of the loop coupling (\ref{hloop});
the change of $N_{12}$ with increasing $M_1$, as described by
eq.~(\ref{binostate}), plays a less important role. The increase of the 
total coupling with increasing $M_1$ nevertheless remains
significant. However, the right panel in Fig.~4 shows that for
$m_{\tilde \chi_1^0} \gsim 15$ GeV this increase of the coupling is
over--compensated by the decrease of the $\beta^3$ threshold factor in
the expression (\ref{hdec}) for the $h$ partial width. \s

Once one--loop corrections are included, for certain values of the
MSSM parameters the branching ratio for invisible $h$ boson decays can
thus reach the level of several percent even if \lsp\ is an almost
purely bino. This would make the detection of these decays possible at
the next generation of $e^+e^-$ linear colliders. At such a collider
it will be possible to isolate $e^+ e^- \rightarrow Z h$ production
followed by $Z \rightarrow \ell^+ \ell^-$ decays ($\ell = e$ or $\mu$)
{\em independent} of the $h$ decay mode, simply by studying the
distribution of the mass recoiling against the $\ell^+ \ell^-$
pair. This allows accurate measurements of the various $h$ decay
branching ratios, including the one for invisible decays, with an
error that is essentially determined by the available statistics
\cite{R11}. Since a collider operating at $\sqrt{s} \sim 300$ to 500
GeV should produce $\sim 10^5$ $Zh$ pairs per year if $|\sin(\alpha -
\beta)| \simeq 1$ one should be able to measure an invisible branching
ratio of about 3\% with a relative statistical uncertainty of about
2\%. \s

\begin{figure}
\begin{center}
\epsfig{figure=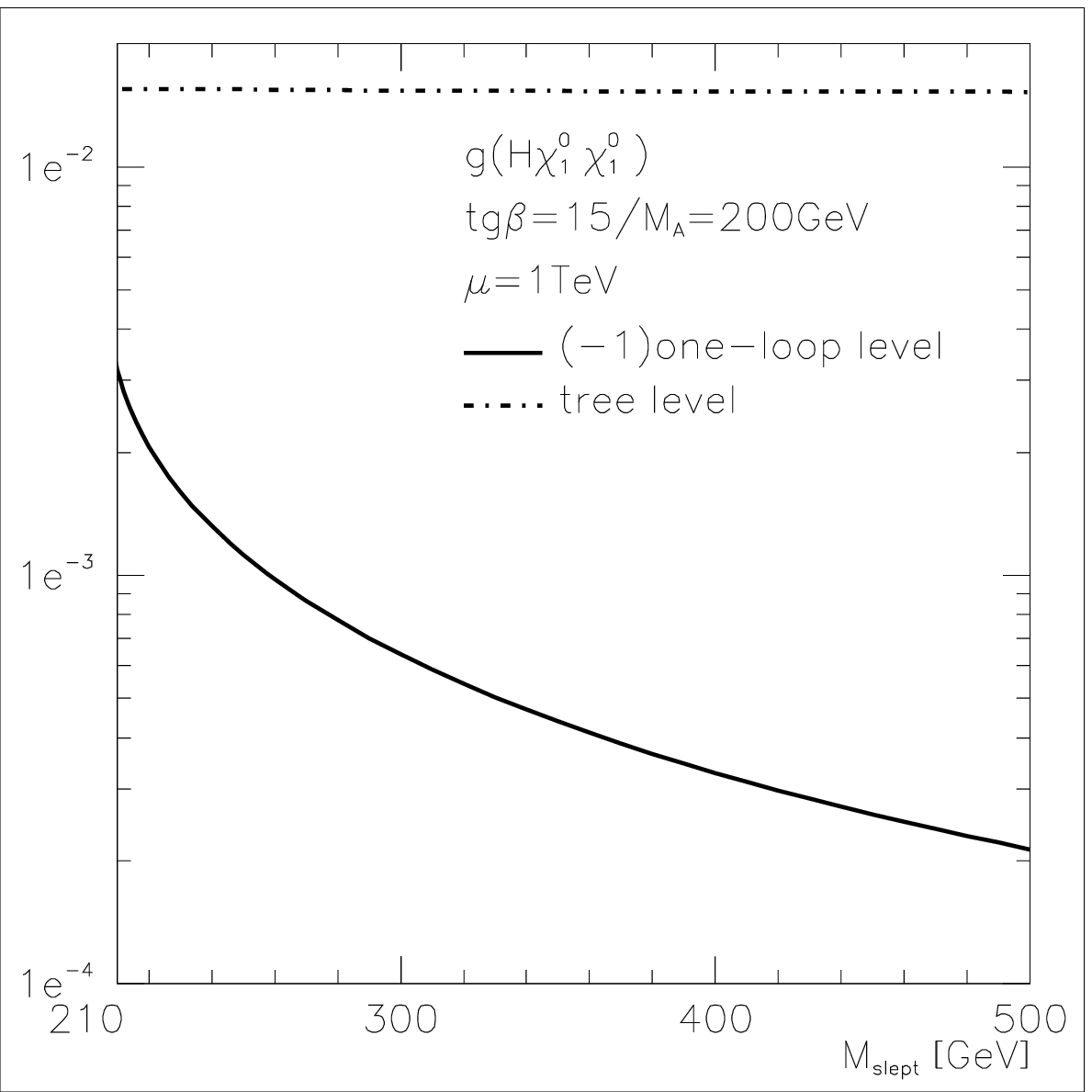,bbllx=2pt,bblly=2,bburx=338,bbury=338,
width=7.5cm,clip=}
\hspace{1.cm}
\epsfig{figure=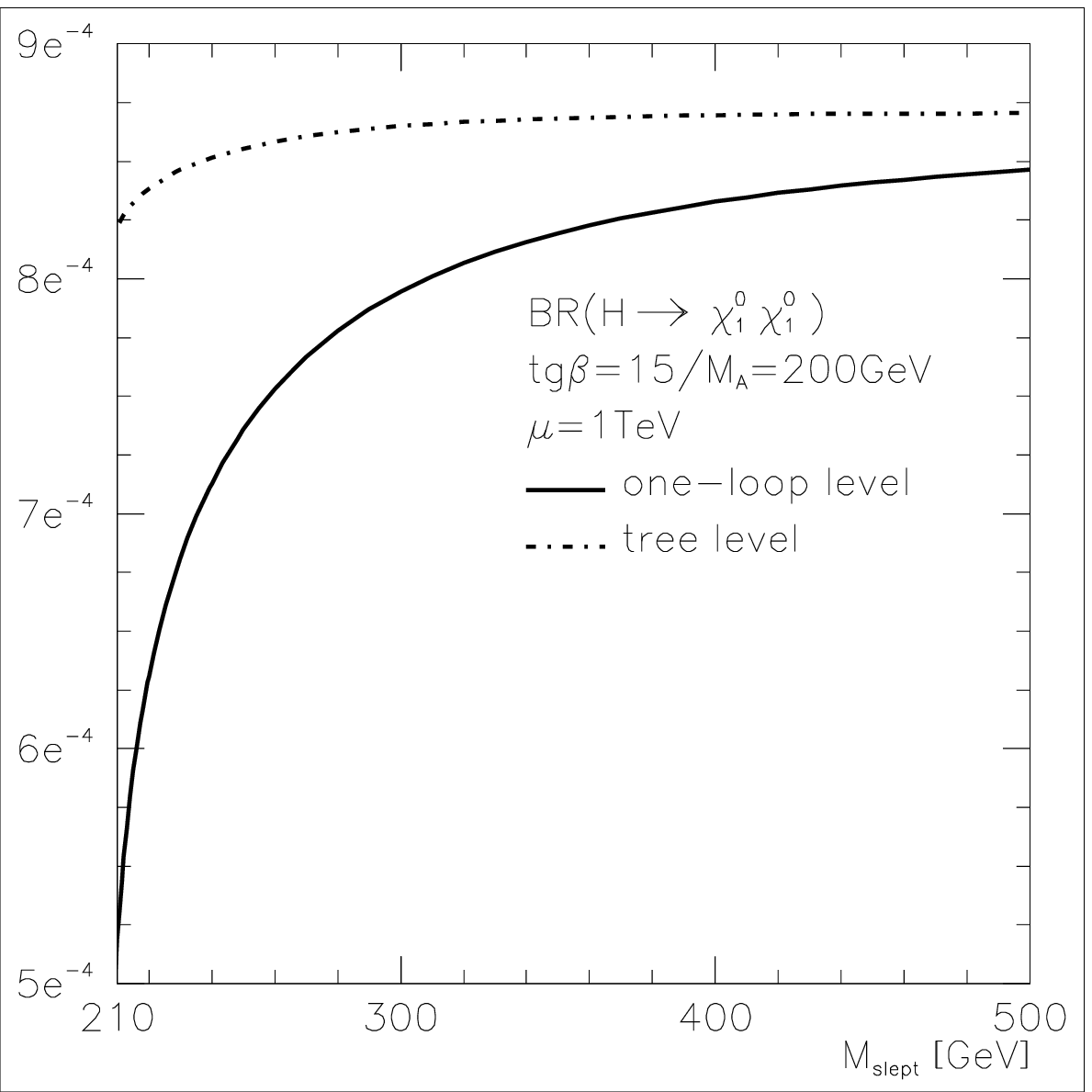,bbllx=2pt,bblly=2,bburx=338,bbury=338,
width=7.5cm,clip=}\\[0.5cm]
\end{center}
\vspace*{1cm}
\begin{center}
\epsfig{figure=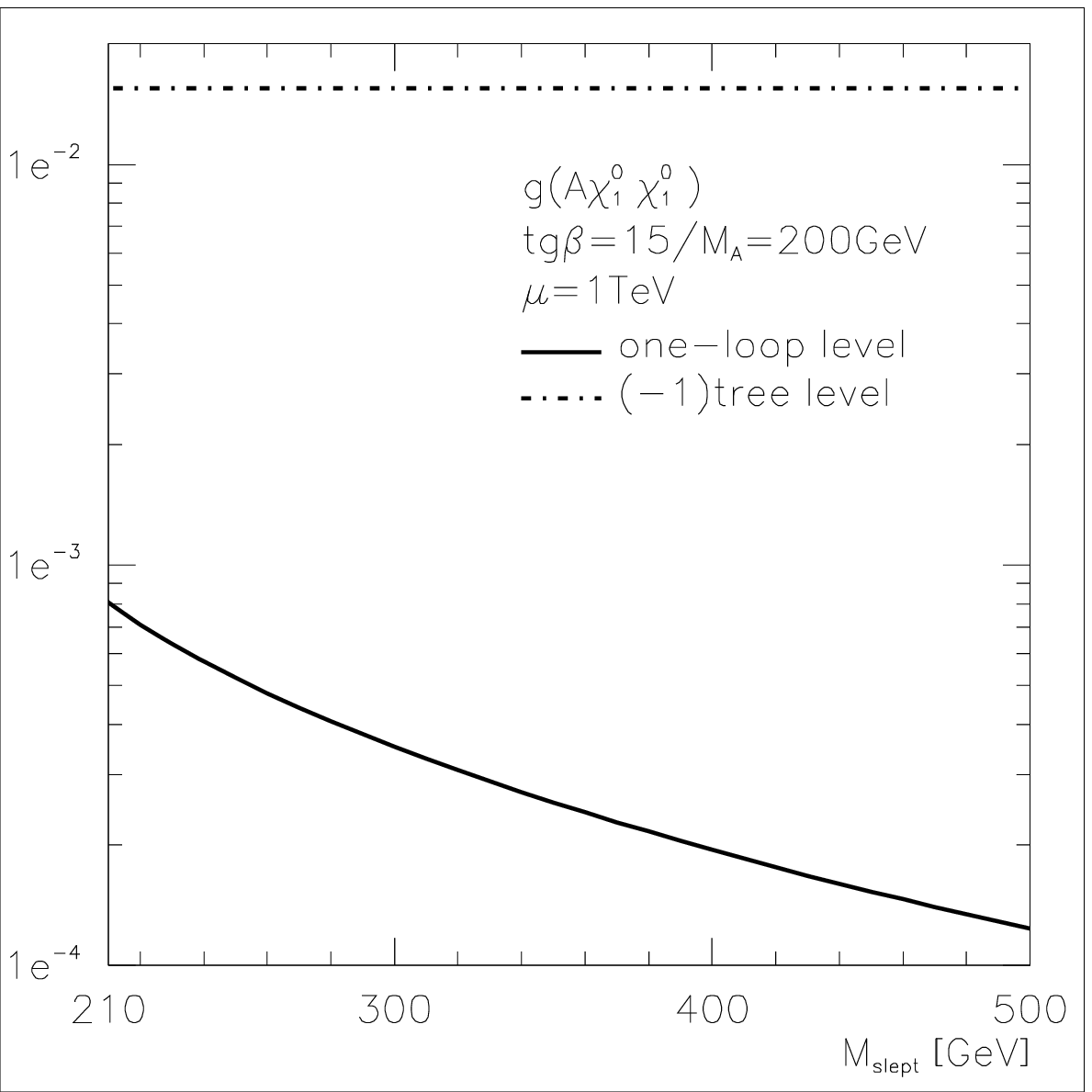,bbllx=2pt,bblly=2,bburx=338,bbury=338,
width=7.5cm,clip=}
\hspace{1.cm}
\epsfig{figure=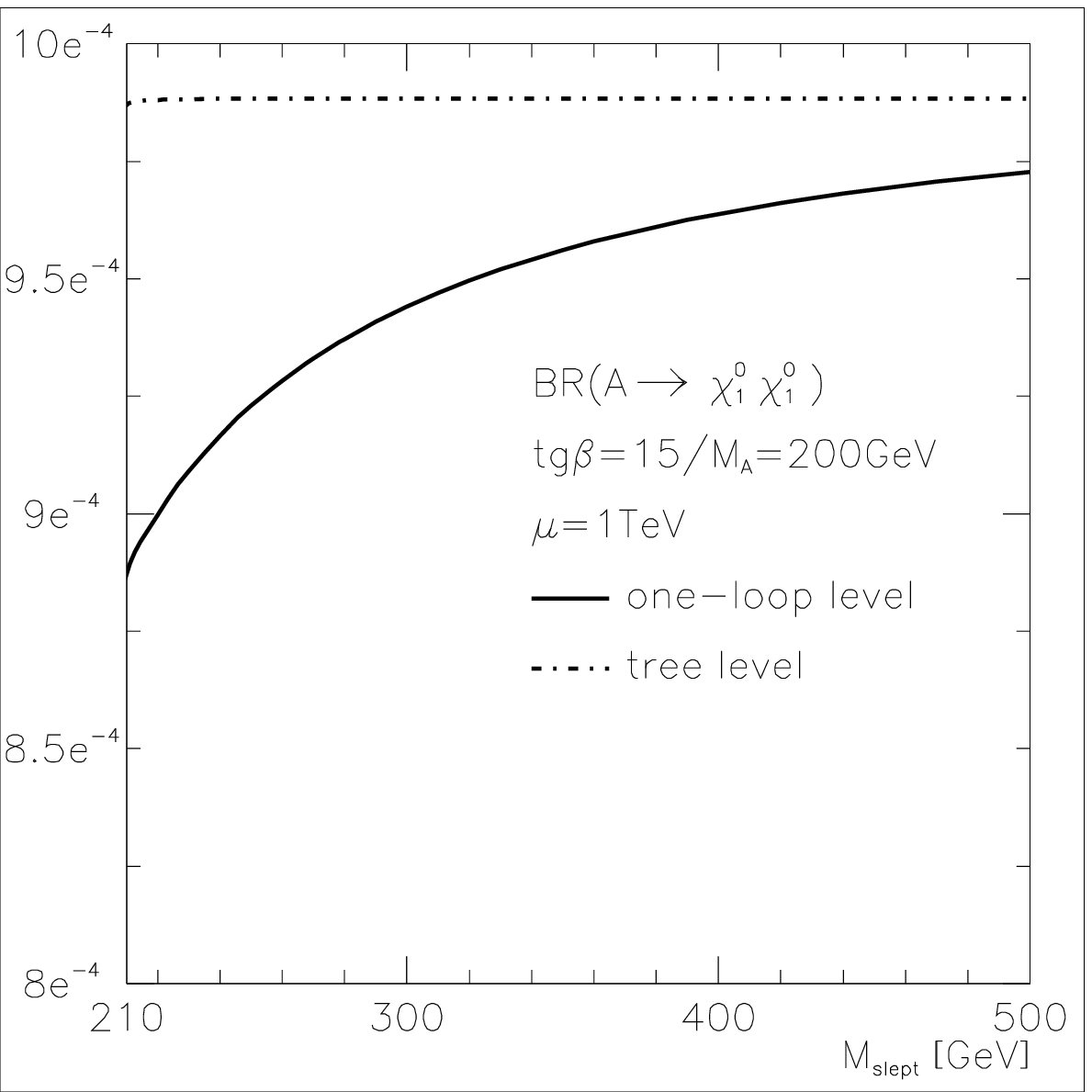,bbllx=2pt,bblly=2,bburx=338,bbury=338,
width=7.5cm,clip=}
\end{center}
\vspace*{.5cm}
\caption{The heavier CP--even Higgs boson $H$ (upper plots) and the 
pseudoscalar $A$ boson (lower plots) couplings (left) and branching ratios 
(right) to pairs of the lightest neutralinos as functions of the common 
slepton mass. The parameters are as in Fig.~2.} 
\end{figure}

We now turn to the heavier MSSM Higgs bosons $H$ and $A$.  The
couplings to the lightest neutralinos are shown in Fig.~5 for the same
input parameters as in Fig.~2. As can be seen, up to a relative minus
sign, the tree--level couplings of these two Higgs bosons are
approximately the same since we are in the decoupling regime where
$d_A \simeq - d_H$, and $e_A \simeq e_H$ with $|e_A| \ll 1$, see
eq.~(\ref{deh}). Eqs.~(\ref{hcoup0}--\ref{deh}) and (\ref{binostate})
also show that the tree--level couplings of the heavy Higgs bosons
exceed that of the light Higgs boson $h$ by a factor $\tan\beta / 2$
[ignoring contributions to eqs.~(\ref{binostate}) with $M_1$ in the
numerator]. On the other hand, the loop corrections are smaller in
case of the heavy Higgs bosons. The corrections to the $H \lsp \lsp$
coupling are reduced by about a factor of 2 compared to the
corrections to the $h \lsp \lsp$ coupling, mostly due to the
relatively smaller coupling to $\tilde t_1$ pairs, see
eqs.~(\ref{hsfsfcoup}). The corrections to the $A \lsp \lsp$ coupling
are even smaller, since the CP--odd Higgs boson $A$ cannot couple to
two identical squarks. The contribution with two $\tilde t_1$ squarks
and one top quark in the loop, which dominates the corrections to the
couplings of the CP--even Higgs bosons for small $m_{\tilde l}$, does
therefore not exist in case of the $A$ boson. As a result, the
corrections to the coupling of $A$ are not only smaller, but also
depend less strongly on $m_{\tilde l}$; recall that for our choice of
parameters $m_{\tilde t_1}$ increases very quickly as $m_{\tilde l}$
is increased from its lowest allowed value of $\sim 210$ GeV, which
comes from the requirement $m_{\tilde t_1} \geq 100$ GeV. Note also
that the $H t t$ and $A t t$ couplings are suppressed by a factor
$\cot\beta$ relative to the $h t t$ coupling, see
eqs.~(\ref{hffcoup}); this becomes important for large squark masses,
where the contributions from Fig.~1b are relatively less important.
Altogether we thus see that the one--loop corrections are much less
important for the heavy Higgs bosons. Note also that for the given set
of parameters they tend to {\em reduce} the absolute size of these
couplings. \s

The partial decay width $\Gamma (H \to
\tilde{\chi}_1^0 \tilde{\chi}_1^0)$ of the CP--even Higgs boson $H$ is
given by eq.~(\ref{hdec}) with $\phi = H$. Due to the different CP
nature of the pseudoscalar Higgs boson $A$, the expression for its
partial decay width differs slightly; it is given by:
\begin{equation} \label{adec}
\Gamma(A \rightarrow \tilde{\chi}_1^0 \tilde{\chi}_1^0) = 
\frac{\beta_A M_A}{16\pi} \left| g^0_{A\tilde{\chi}_1^0\tilde{\chi}_1^0}
 + g^1_{A\tilde{\chi}_1^0\tilde{\chi}_1^0} \right|^2 .
\end{equation}
Again, because in the decoupling regime the CP--even $H$ boson and the
pseudoscalar $A$ boson have almost the same couplings to Standard
Model particles and to the neutralinos [at the tree level], their
branching ratios are approximately the same. The one--loop
contributions decrease the branching ratios by at most $\sim 10$ to
40\%. Note that the total decay widths of the $A$ and $H$ bosons are
strongly enhanced by $\tan^2 \beta$ factors [$\Gamma (H, A \ra
b\bar{b}) \propto m_b^2 g_{A,Hbb}^2$]. This over--compensates the
increase of their couplings to neutralinos, so that their branching
ratios into \lsp\ pairs are far smaller than that of the light Higgs
boson $h$, remaining below the 1 permille level over the entire
parameter range shown. Moreover, the cross section for the production
of heavy Higgs bosons at $e^+e^-$ colliders is dominated by associated
$HA$ production, which has a much less clean signature than $Z h$
production does. Branching ratios of the size shown in Fig.~5 will
therefore not be measurable at $e^+e^-$ colliders. In fact, they will
probably even be difficult to measure at a $\mu^+ \mu^-$ collider
``Higgs factory''; recall that the $Z$ factories LEP and SLC ``only''
determined the invisible decay width of the $Z$ boson to $\sim 0.1\%$,
see eq.~(\ref{zexp}).

\section{SUSY Dark Matter}

It is well known that in the MSSM with exact R-parity, the lightest
neutralino is a good cold Dark Matter (DM) candidate
\cite{NeutDM,SDM}. For a very reasonable range of supersymmetric
parameters the relic density of a bino--like neutralino satisfies $0.1
\lsim \Omega_{\tilde \chi_1^0} h^2 \lsim 0.3$, which is the currently
preferred range \cite{omega}; here $\Omega$ denotes the mass density
in units of the critical density, and $h$ is today's Hubble constant
in units of 100 km/(sec$\cdot$Mpc). Note that this argument singles
out a bino--like LSP. A higgsino-- or wino--like LSP would have a
thermal relic density in this range only if its mass is around 1 TeV;
such a large mass for the {\em lightest} superparticle would be
difficult to reconcile with finetuning or naturalness arguments. \s

The two relic neutralino search strategies that suffer least from our
lack of knowledge of the DM distribution throughout our galaxy are
``direct'' detection, where one looks for the elastic scattering of
ambient neutralinos off nuclei in a detector; and the ``indirect''
search for high--energy neutrinos originating from \lsp \lsp\
annihilation in the center of the Earth or Sun. In both cases the
signal rate is directly proportional to the LSP--nucleon cross section
\cite{SDM}. This cross section in turn depends on the
neutralino--quark interaction strength, on the distribution of quarks
in the nucleon, and on the distribution of nucleons in the nucleus. \bigskip
 
\begin{picture}(1000,170)(10,0)
\ArrowLine(80,150)(110,120)
\ArrowLine(80,90)(110,120)
\Text(140,130)[]{$Z, \Phi$}
\DashArrowLine(110,120)(160,120){4}{}
\ArrowLine(160,120)(190,150)
\ArrowLine(160,120)(190,90)
\Text(70,155)[]{$\tilde{\chi}_1^0$}
\Text(70,85)[]{$\tilde{\chi}_1^0$}
\Text(195,155)[]{$q$}
\Text(195,85)[]{$q$}
\ArrowLine(280,150)(310,120)
\ArrowLine(280,90)(310,120)
\Text(340,130)[]{$\tilde{q}$}
\DashArrowLine(310,120)(360,120){4}{}
\ArrowLine(360,120)(390,150)
\ArrowLine(360,120)(390,90)
\Text(270,155)[]{$\tilde{\chi}_1^0$}
\Text(395,155)[]{$\tilde{\chi}_1^0$}
\Text(270,85)[]{$q$}
\Text(395,85)[]{$q$}
\end{picture}
\vspace*{-2.3cm}

\nn \centerline{Figure 6: Diagrams contributing to the effective 
neutralino--quark interactions.}
\vspace*{4mm} 
\setcounter{figure}{6}

Three classes of diagrams contribute to the neutralino quark
interaction: the exchange of a $Z$ or Higgs boson in the $t-$channel,
and squark exchange in the $s-$ or $u-$channel, see
Fig.~6. $Z$ exchange only leads to a spin--dependent
interaction, while Higgs exchange contributes only to the
spin-independent interaction, and squark exchange gives both
spin--dependent and spin--independent contributions, where the latter
arises only due to combinations of couplings that violate chirality;
see Ref.~\cite{SDM} for details. In general the spin--independent
contribution is more important since it leads to a {\em coherent}
coupling to heavy nuclei\footnote{The \lsp\ capture rate in the Sun,
which contains very few heavy nuclei, can also get important
contributions from spin--dependent interactions.}, and the Higgs
exchange contribution is usually bigger than the squark exchange
contribution. \s

We saw in Sec.~2.1 that in the pure gaugino limit the $\lsp\lsp$Higgs
couplings vanish at the tree level. We just stated that the most
important contribution to the spin--independent neutralino quark
interaction often comes from Higgs exchange. In this Section we
therefore study the effect of the fermion--sfermion loop corrections
to the Higgs couplings to neutralinos on the neutralino--proton cross
section. We use the expressions in Ref.~\cite{MD3}; these expressions
also include contributions from effective LSP--gluon interactions. We
also take into account leading SUSY QCD corrections to the scattering
cross section \cite{DD}, and assume the ``standard'' value \cite{SDM}
for the strange contribution to the nucleon mass, $m_s \langle p |
\bar s s | p \rangle = 130$ MeV. \s

\begin{figure}[htbp]
\begin{center}
\epsfig{figure=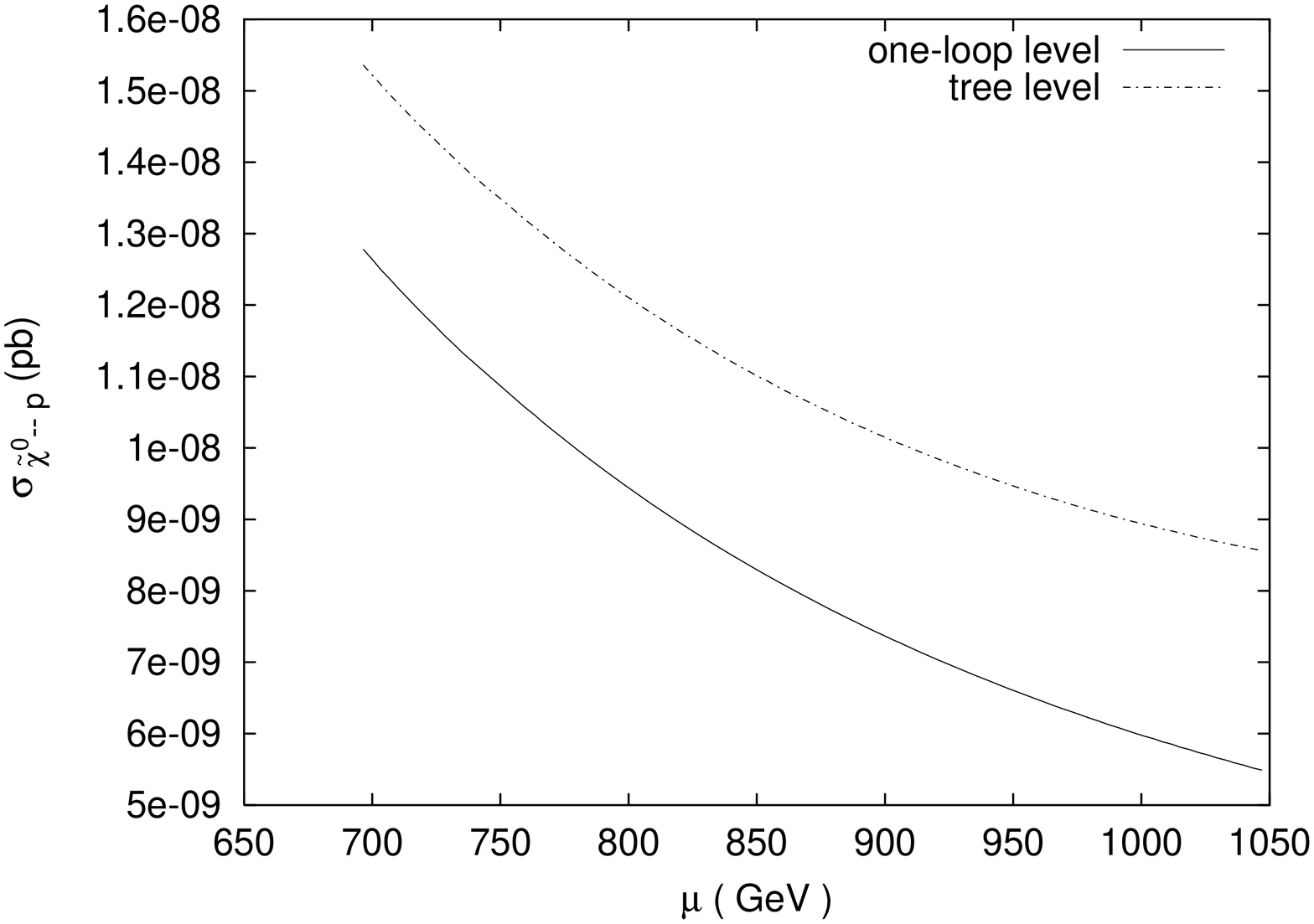,angle=-90,width=0.8\textwidth,clip=}
\vspace*{1.5cm}
\epsfig{figure=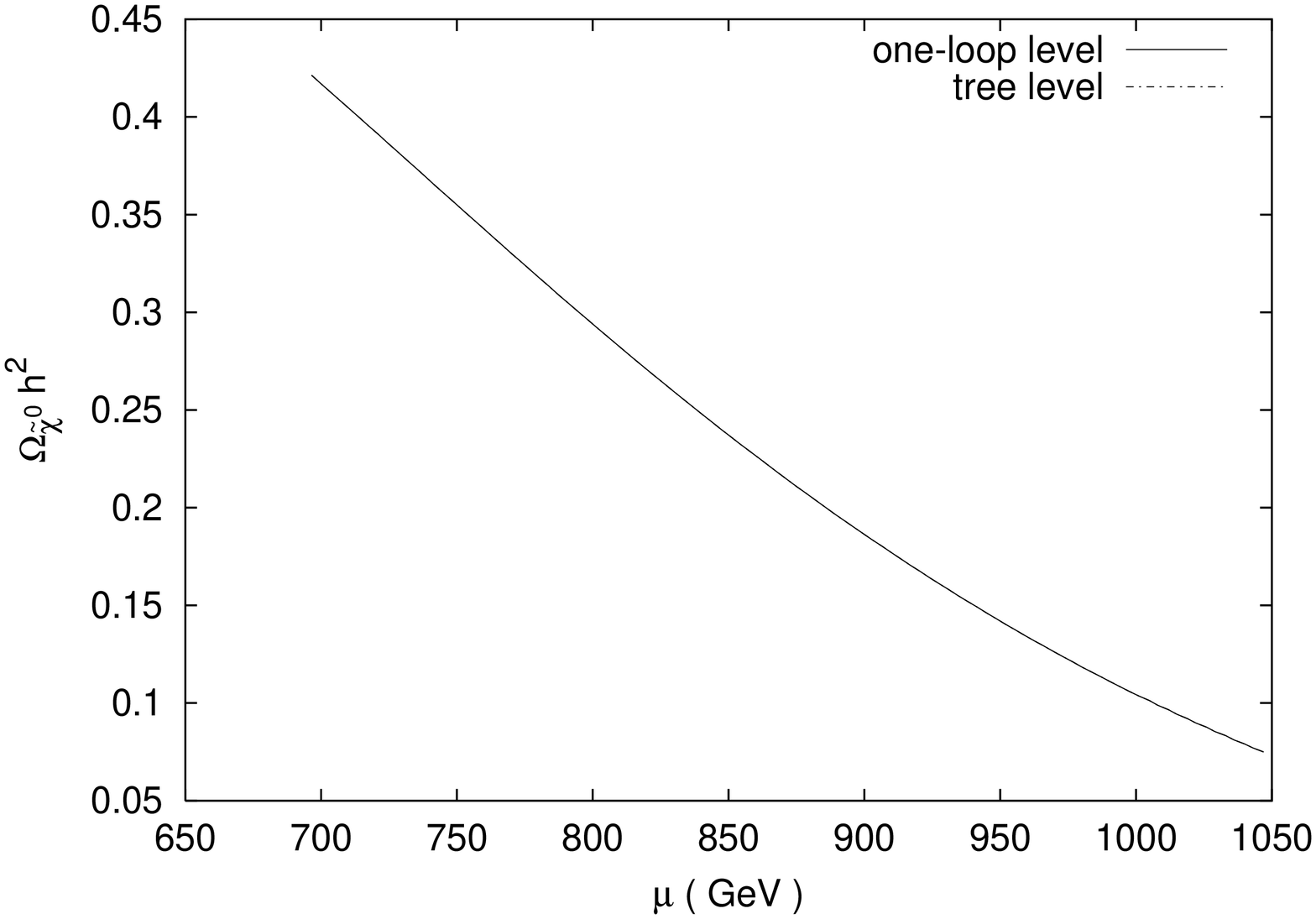,angle=-90,width=0.8\textwidth,clip=}
\end{center}
\vspace*{-1.5cm}
\caption{The predicted neutralino--proton scattering cross section
and the thermal relic density (bottom), when
one--loop corrections to the couplings of the Higgs boson to the LSP
are included (solid curves) or omitted (dashed curves); in the lower
frame the two curves practically lie on top of each other. The values
of the relevant parameters are: $m_{\tilde\chi_1^0} \simeq M_1= 50$
GeV, $m_{\tilde q} = 400$ GeV, $M_A = 200$ GeV, $M_2 = 105$ GeV, $A_q
= 2.73 m_{\tilde q}, \ \tan\beta = 21, \ m_{\tilde e_R} = 210$ GeV,
and $m_{\tilde \nu} = 200$ GeV.}
\label{r50}
\end{figure}

As in Sec.~2 we use non--universal soft breaking terms but keep all
squark masses identical. All experimental constraints on sparticle and
Higgs boson masses are taken into account. We show two illustrative
examples. In Fig.~7 we take $m_{\tilde\chi_1^0} \simeq M_1 = 50$ GeV,
and vary $\mu$ from 700 to 1050 GeV. The values of the other
parameters are similar to those of Fig.~2. The upper panel shows the
predicted relic LSP proton scattering cross section, while the lower
panel shows the thermal LSP relic density $\Omega_{\tilde \chi_1^0}
h^2$. In the case at hand this latter quantity is essentially
determined \cite{SDM} by \lsp\ pair annihilation into charged lepton
pairs. The rapid decrease with increasing $\mu$ is due to the decrease
of $m_{\tilde \tau_1}$, which is caused by enhanced $\tilde \tau_L -
\tilde \tau_R$ mixing; this not only trivially increases the $\tilde
\tau_1$ propagator, but also allows \lsp\ annihilation to proceed from
an S--wave initial state \cite{MD1}. We include this figure here in
order to illustrate that our choice of parameters leads to an LSP
relic density of the required magnitude. Of course, this is not
difficult to arrange, since the loop corrections we are interested in
are almost independent of soft breaking parameters in the slepton
sector, which therefore always can be chosen to produce the desired
relic density, as long as the LSP is not too heavy. Not surprisingly,
the prediction of $\Omega_{\tilde \chi_1^0} h^2$ is practically not
affected by the loop corrections presented in Sec.~2. \s

On the other hand the effect of these corrections on the LSP--nucleon
scattering cross section can be significant, as shown in the upper
frame of Fig.~7, where they reduce the predicted scattering rate by up
to a factor of 1.5. For the given choice of parameters the tree--level
and one--loop contributions to the $h \lsp \lsp$ coupling have
opposite sign, as in the upper frames of Fig.~3; the absolute value of
the sum of these contributions only amounts to about 20\% of the
absolute value of the tree--level coupling. However, the scattering
cross section is actually dominated by the exchange of the heavier $H$
boson. For the given choice of parameters its mass exceeds $M_h$ only
by a factor of $\sim 1.5$, and its couplings to down--type quarks are
enhanced by a factor $\sim \tan \beta$; moreover, we saw in Sec.~3
that $\left| g^0_{H \tilde \chi_1^0 \tilde \chi_1^0} \right|$ exceeds
$\left| g^0_{h \tilde \chi_1^0 \tilde \chi_1^0} \right|$ by a factor
$\sim \tan\beta / 2$. The absolute value of the $H \lsp \lsp$ coupling
is also reduced here, as in Fig.~5, but only by $\sim 10\%$. Another
significant contribution comes from $\tilde s$ squark exchange, since
$\tilde s_L - \tilde s_R$ mixing is enhanced for large $|\mu| \cdot
\tan\beta$; this mixing increases the spin--independent part of the
squark exchange contribution \cite{sredwat}. Of course this latter
contribution, which interferes constructively with the $H-$exchange
contribution, is not affected by the loop corrections of Sec.~2. Hence
in the given case the net effect of these corrections is significantly
smaller than in case of the invisible width of the $h$ boson. Finally,
the overall decline of the predicted LSP scattering rate with
increasing $\mu$ is due to the reduction of the higgsino components of
the LSP, see eqs.~(\ref{binostate}), which leads to smaller tree--level
LSP couplings to the Higgs bosons. Since the one--loop corrections to
the Higgs couplings depend only rather weakly on $\mu$ they become
relatively more important when $|\mu|$ is increased. \s

As a second example, shown in Fig.~8, we take $m_{\tilde \chi^0_1}
\simeq M_1 = 100$ GeV, and slightly reduced values for $m_A$ and
$\tan\beta$. The reduction of $\tan\beta$ implies less $\tilde \tau_L
\tilde \tau_R$ mixing, so that the relic density is much less
sensitive to $\mu$ than in the previous example. On the other hand,
the increase of the LSP mass while keeping the slepton masses
essentially the same leads to an increase of the annihilation cross
section, so that we now obtain a relic density of the required size in
the entire range of $\mu$ shown in this figure. The effect of the loop
corrections to the Higgs \lsp\lsp\ vertices is somewhat larger here
than in the previous example. Due to the reduced value of $M_A$ we are
now in a regime with strong Higgs mixing, i.e. $|\sin \alpha|$ is much
larger than the value $\cos\beta$ it takes in the limit $M_A \ra
\infty$. As a result the exchange of the lighter $h$ boson is now more
important than before. The loop corrections reduce the absolute value
of the $h \lsp \lsp$ coupling by about a factor of two, which leads to
a similar reduction of the predicted LSP scattering rate.   \s

\begin{figure}
\begin{center}
\epsfig{figure=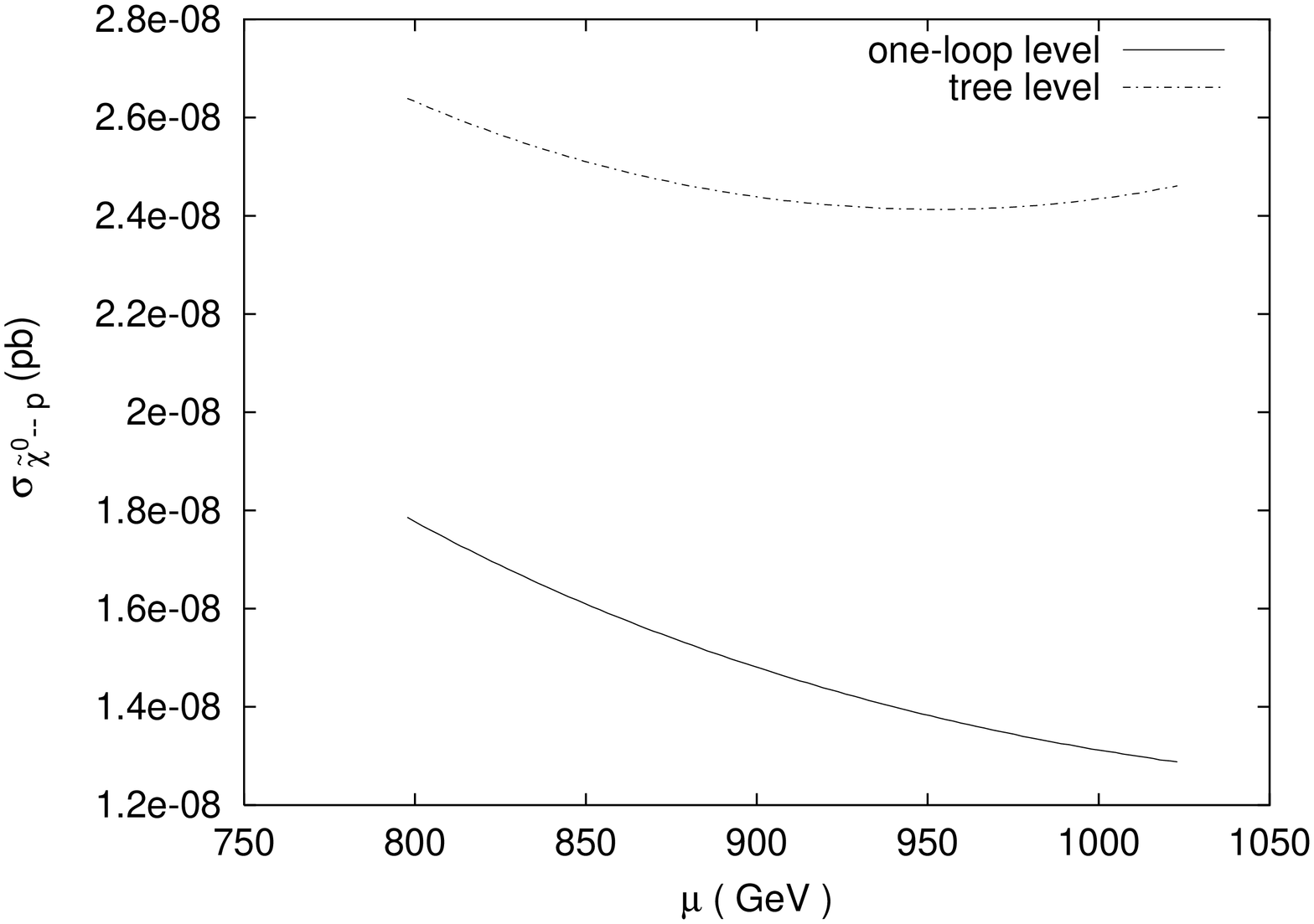,angle=-90,width=0.8\textwidth,clip=}
\vspace{1.5cm}
\epsfig{figure=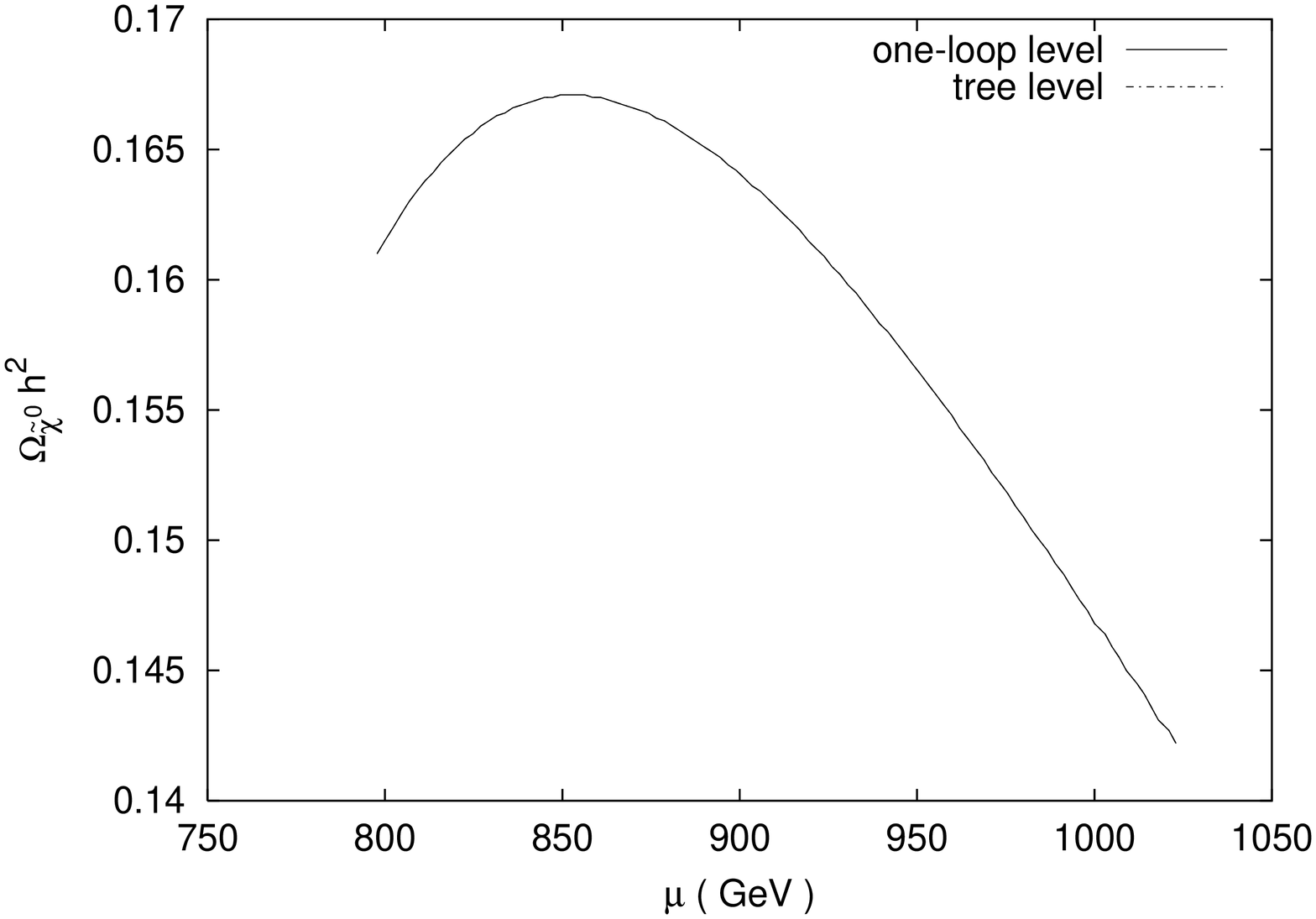,angle=-90,width=0.8\textwidth,clip=}
\end{center}
\vspace*{-1.5cm}
\caption{As in Figure 7, but for the following set of input parameters:
 $m_{\tilde \chi_1^0} = 100$ GeV, $m_{\tilde q} = 400$ GeV, $M_A = 150$
GeV, $M_2 = 200$ GeV, $A_q = 2.77 m_{\tilde q}, \ \tan\beta = 14$, and
$m_{\tilde e_R} =  m_{\tilde \nu} = 200$ GeV.}
\label{r100}
\end{figure}

\begin{figure}
\begin{center}
\epsfig{figure=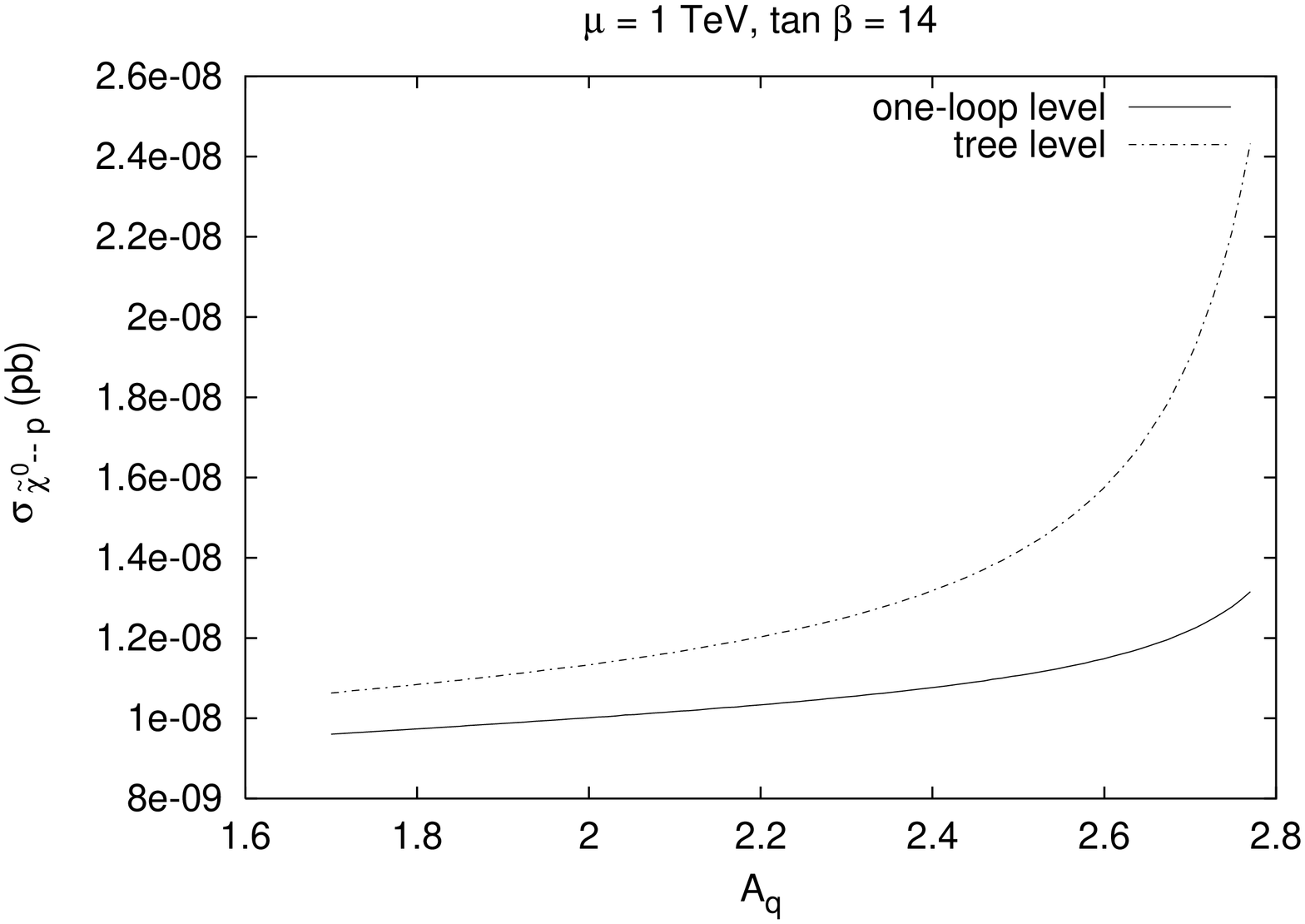,angle=-90,width=0.8\textwidth,clip=}
\vspace{1.5cm}
\epsfig{figure=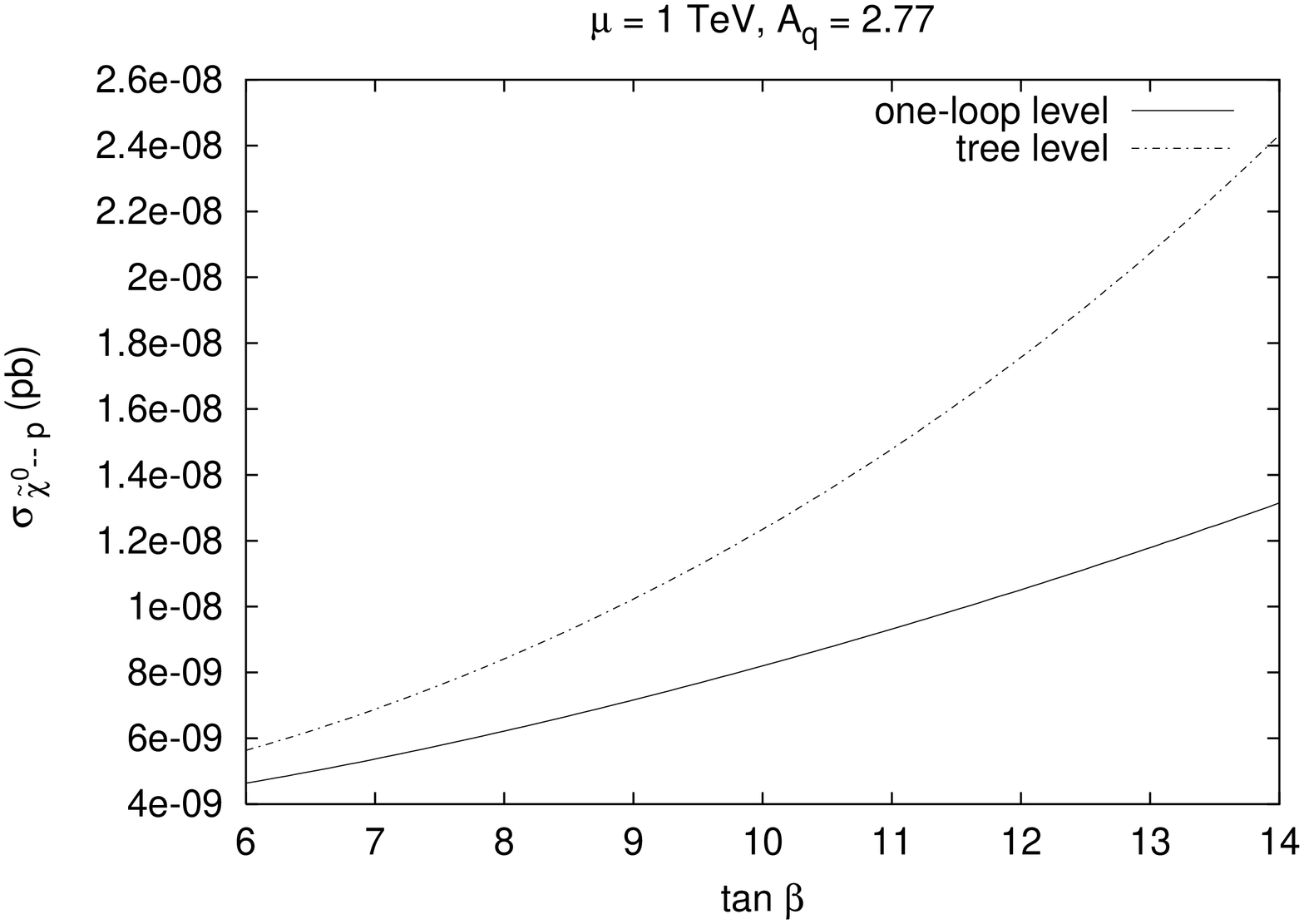,angle=-90,width=0.8\textwidth,clip=}
\end{center}
\vspace*{-1.5cm}
\caption{The dependence of the $\lsp-p$ cross section on $A_t$ (top)
and $\tan\beta$ (bottom) for $\mu = 1$ TeV. The values of the other
parameters are as in Fig.~8. The upper bound on $A_t$ and both the
upper and lower bound on $\tan\beta$ are determined by the
experimental lower bound on $m_h$.}
\label{r100a}
\end{figure}

We already saw in Sec.~3.2 that loop corrections to the Higgs\lsp\lsp\
couplings are maximal if $\tilde t_1$ is light and the Higgs$\tilde
t_1 \tilde t_1$ couplings are large. This is again illustrated in
Fig.~9, where we show the dependence of $\sigma_{\lsp-p}$ on $A_t$
(top) and $\tan\beta$ (bottom). Reducing $A_t$ from its upper bound
(set by the experimental lower bound on $m_h$) increases $m_{\tilde
t_1}$ and reduces the $h \tilde t_1 \tilde t_1$ coupling, which leads
to a rapid decrease of the loop correction to $\sigma_{\lsp-p}$. The
$A_t$ dependence of the tree--level prediction is due to the $t -
\tilde t$ loop corrections to the MSSM Higgs sector, which are always
included. This dependence is quite mild, except near the upper bound
on $A_t$ where both $m_h$ and the mixing angle $\alpha$ depend
sensitively on $A_t$. On the other hand, the tree--level prediction
for the scattering cross section depends quite strongly on
$\tan\beta$, due to the $1/\cos\beta-$behavior of the Yukawa couplings
of down--type quarks. However, the negative loop corrections also
increase in size with increasing $\tan\beta$, largely due to the
$\tan\beta$ dependence of the Higgs couplings to $t$ and $\tilde t_1$
(recall that we are far away from the ``decoupling limit''). We thus
see that in the case at hand the parameter dependence actually becomes
weaker once loop corrections are included; however, this is not always
the case. Finally, we note that scenarios with $m_{\tilde t_1} \sim
m_h$, where our loop corrections will be sizable, can also be found
\cite{kon} in models with universal soft breaking terms at the scale
of Grand Unification, if the trilinear soft breaking parameter is
sizable already at this scale; the condition $|\mu| \gg M_1$ is almost
always satisfied in these models. \s

In these two examples the corrections to the Higgs couplings to the
LSP reduce the predicted scattering rate. It is easy to construct
scenarios with even larger but positive corrections, by taking very
large values for first and second generation squark masses as well as
for $M_A$, while keeping third generation squark masses relatively
small; this kind of spectrum is e.g. expected in ``more minimal''
supersymmetric models \cite{mms}.  The largest contribution to the
LSP--nucleon scattering cross section then comes from $h$ exchange, so
the corrections to this cross section would scale essentially like the
square of the correction to the $h \lsp \lsp$ coupling discussed in
Sec.~2.2. However, the predicted counting rate in such a scenario
would be very low, well below the sensitivity of near future
experiments. We finally note that the corrections to the $Z \lsp \lsp$
couplings will have some effect on the spin--dependent LSP--nucleon
scattering cross section. However, this cross section is usually
dominated by first and second generation squark exchange, which in
this case does not require violation of chirality. We therefore expect
the loop corrections to the $Z$ coupling to be significant only in
scenarios where $|M_1| \ll |\mu| \lsim m_{\tilde u, \tilde d, \tilde
s}$ but some sfermion masses are significantly smaller than $|\mu|$. \s

It should be noted that currently the $\lsp p$ scattering cross
section can only be predicted to within a factor of 2 or so even if
the fundamental LSP--quark interactions were known exactly, the main
uncertainty coming from the poorly known size of the strange matrix
element $m_s \langle p | \bar s s | p \rangle$. For the cases
considered in this paper, the loop corrections calculated here can
therefore shift the cross section by at most one ``theoretical
standard deviation''. However, this theoretical uncertainty will
presumably be reduced in future. We believe that the calculation
presented here, together with the results of refs.\cite{higgsino} and
\cite{DD}, reduces the theoretical uncertainty of the prediction of
the hard scattering cross section for given SUSY parameters to the
level of 10\%.

\section{Summary and conclusions}

In this paper we have studied quantum corrections to the couplings of
the $Z$ and neutral Higgs bosons to gaugino--like neutralinos in the
MSSM. We focused on the phenomenologically most interesting case of a
bino--like lightest neutralino \lsp\ as LSP, but our analytical
results of Sec.~2 are valid for a more general gaugino--like
neutralino, irrespective of whether it is the LSP. We found that these
corrections can completely dominate the tree--level contribution to
the coupling of the lightest CP--even Higgs boson. The corrections to
the couplings of the $Z$ and heavy CP--even Higgs boson are somewhat
less significant, but can still amount to about a factor of 2. Since the
CP--odd Higgs boson cannot couple to two identical sfermions the
corrections are suppressed in this case. In all cases the corrections
can be significant only if some sfermion masses are considerably
smaller than the supersymmetric higgsino mass $|\mu|$. Both (s)lepton
and (s)top loop contributions to the $Z$ coupling can be important,
since the experimental bounds for $m_{\tilde t_1}$ and $m_{\tilde l}$
are still quite close to $M_Z$. The Higgs couplings receive their
potentially largest corrections from loops involving third generation
quarks and their superpartners; in the latter case the corrections are
also quite sensitive to the size of the trilinear soft breaking
parameter $A_t$ [and $A_b$, if $\tan\beta \gg 1$]. \s

Turning to applications of these calculations, we found that these
corrections might change the predicted detection rate of Dark Matter
LSPs by up to a factor of two even for scenarios where the rate is
close to the sensitivity of the next round of direct Dark Matter
detection experiments. The possible impact on the invisible width of
the lightest CP--even Higgs boson $h$ is even more dramatic: it could
be enhanced to a level that should be easily measurable at future
high--energy $e^+ e^-$ colliders, even if the LSP is an almost perfect
bino. This would open a new window for testing the MSSM at the quantum
level.

\subsection*{Acknowledgments}
This work is supported in part by the Euro--GDR Supersym\'etrie and by
the European Union under contract HPRN-CT-2000-00149. MD thanks the
KIAS school of physics in Seoul, Korea for their hospitality during
the completion of this work; he is partly supported by the Deutsche
Forschungsgemeinschaft through the SFB 375. PFP thanks the ICTP,
Trieste, for hospitality while this work was completed.

\end{document}